\documentclass[useAMS]{mn2e}
\usepackage{amssymb,amsmath}
\usepackage{graphicx}

\title[Shifted Broad Emission Lines]
{partly obscured accretion disk model to explain Shifted broad Balmer emission lines of Active Galactic Nuclei}

\author[Zhang et al.]
       {Xue-Guang Zhang$^{1}$\thanks{xueguang@MPA-Garching.MPG.De},
        Deborah Dultzin$^2$,
        Ting-Gui Wang$^3$,
        Guinevere Kaufmann$^{1}$\\
       $^1$Max-Plank Institute f\"ur Astrophysik,
           Karl-Schwarzschild-Strasse 1, 85748 Garching, Germany  \\
       $^2$Instituto de Astronomia, Universidad Nacional Autonoma de
                 Mexico, Apdo Postal 70-264, Mexico D. F. 04510, Mexico \\
       $^3$Center for Astrophysics, Department of astronomy and Applied
                 Physics, University of Science and Technology of China, \\
                 Hefei, Anhui, P.R.China}

\date{}
\pagerange{\pageref{firstpage}--\pageref{lastpage}} \pubyear{2008}
\def\LaTeX{L\kern-.36em\raise.3ex\hbox{a}\kern-.15em
    T\kern-.1667em\lower.7ex\hbox{E}\kern-.125emX}

\begin{document}
\label{firstpage}

\maketitle

\begin{abstract}
We present a new model to explain the appearance of red/blue-shifted 
broad low-ionization emission lines, especially emission lines in optical 
band, which is commonly considered as an indicator of radial motion of the  
line emitting gas in broad emission line regions (BLRs) of Active Galactic 
Nuclei (AGN). We show that partly obscured disk-like BLRs of dbp emitters 
(AGN with double-peak broad low-ionization emission lines) can also 
successfully produce shifted standard Gaussian broad balmer emission lines. 
Then we select eight high quality objects (S/N gt 40 at r band) with shifted 
standard gaussian broad H$\alpha$ (the shifted velocity larger than 
180 ${\rm km\cdot s^{-1}}$) from SDSS DR4. All eight selected objects have 
visible stellar absorption features in their spectra, except SDSS J1007+1246, 
which allows us to estimate BH masses through M-sigma relation which 
has proven to be the most reliable method. We also calculate virial BH masses 
from continuum luminosity and line width of broad H$\alpha$, assuming broad 
emission line from "normal" BLRs dominated by virialized motions. We find that 
the BH masses calculated from M-sigma relation are systematically larger 
than virial BH masses for the selected objects, even after the correction 
of internal reddening effects in BLRs. The smaller virial BH masses than BH 
masses from M-sigma relation for objects with shifted broad emission lines 
are coincident with what we expect from the partly obscured accretion disk model. 
Thus, we provide an optional better model to explain the appearance of shifted 
broad emission lines, especially for those objects with underestimated 
virial BH masses. Finally, we make predictions about the variation of shifted 
broad H$\alpha$ with  time passage for the two models: Broad H$\alpha$ from "normal" radially moving clouds or broad H$\alpha$ from dbp BLRs in a partly obscured accretion disk.
\end{abstract}

\begin{keywords}
Galaxies:Active -- Galaxies:Nuclei -- quasars:Emission lines
\end{keywords}

\section{Introduction}

   There is no way to spatially resolve broad emission line regions (BLRs) of 
active galactic nuclei (AGN) by direct observation, and no hope of doing so in
the foreseeable future. The information about geometry and kinematics/dynamics 
of BLRs closer to central black hole (BH) can be obtained from study of 
properties of broad emission lines in observed spectra of AGN. In order to 
carry out this type of analysis, high signal to noise (S/N gt 30) spectra are required. Pioneer work on broad line profiles based on high S/N quasar spectra 
prior to the SDSS database (Sloan Digital Sky Survey, Adelman-McCarthy et al. 
2006) is summarized in Sulentic et al. 2000. An atlas of more than 200 broad H$\alpha$ (some with higher S/N than data from SDSS) can be found in 
Marziani et al. (2003).  Broad emission lines can be roughly classified into 
two kinds according to line profiles. Most of them have approximately 
logarithmic forms, sometimes with slight asymmetry. A small part of them have 
very complex profiles, such as the double-peaked line profiles. Besides line 
profiles, variations of broad emission lines, especially variations of strength, 
have been studied for several decades. From the variations of the broad lines 
as a response to variations in ionizing continuum emission, the size of BLRs 
(here 'size' means the distance between central region and BLRs) can be 
estimated by the reverberation mapping technique (Peterson 1993, 2001) based 
on pioneer theoretical work by Blandford \& McKee (1982). In practice, there are 
some problems with the application of this technique, e.g. Maoz (1996) has 
shown that there is not unique one-dimensional transfer function (which can 
provide some information about geometry of BLRs) for NGC5548. Not to mention 
the observational effort to monitor the variability of both lines and 
continuum simultaneously, this has been accomplished for no more than 
50 low redshift objects (Collin 2007).

   Both from reverberation mapping technique and analysis of high S/N line 
profiles, we have been able to gather some information on velocity fields of 
BLRs of AGN. It is widely accepted that the profiles clearly have Doppler 
origin, and commonly assumed that BLRs clouds are mainly gravitationally 
dominated by central mass of host galaxy (Gaskell 1996, 1988; Wandel et al. 
1999; Peterson \& Wandel 1999). However, the basic assumptions can not explain 
asymmetry in part of broad emission lines. Although there are several 
possible explanations for the asymmetry, the first and most 
obvious one is that there is radial motion of BLRs clouds , which 
has been proved by cross-correlation test (Netzer 1990). Thus asymmetry 
in broad emission lines indicates that part of emission line clouds in BLRs 
can be moving away, or perhaps into the central region of AGN.  Obviously, 
if radial flow in BLRs is the dominant component, one shifted logarithmic 
profile should be expected. Recently, Bonning et al. (2007) 
proposed to find recoiling black holes system in SDSS through shifted broad 
emission lines due to radial motion in BLRS, although no convincing 
evidence to prove recoiling BH in AGN. 

   Besides normal AGN with broad emission lines having logarithmic form, 
there is one special kind of AGN, AGN with {\bf D}ou{\bf B}le-{\bf P}eaked 
low-ionization broad emission lines (hereafter, dbp emitters), which have 
undoubtfully distinct properties of broad emission lines. There are 
three famous dbp emitters: NGC 1097(Storchi-Bergmann, Nemmen, et al. 2003,  
Storchi-Bergmann, Eracleous et al. 1997, Storchi-Bergmann, Eracleous \& Halpern  
1995, Storchi-Bergmann, Baldwin et al. 1993), Arp102B (Chen et al. 1989a,
1989b, 1997, Halpern et al. 1996, Antonucci et al. 1996, Sulentic et al.
1990) and 3C390.3 (Shapovalova et al. 2001, Gilbert et al. 1999). There
are at least two different models proposed to explain observed features of 
double-peaked broad emission lines. One is the binary black hole model 
(Begelman et al.1980, Gaskell 1983). However this model has failed to account
for properties of long-term variability and leads to much
larger central BH masses than $10^{10}{\rm M_{\odot}}$ (Eracleous et al. 1997).
Thus, we prefer to the other model, the accretion disk model. This model was 
first proposed by Chen et al. in 1989 (Chen et al. 1989a, 1989b), and then 
improved from circular accretion disk model to elliptical accretion disk 
model by Eracleous et al. (1995). The accretion disk model can be successfully 
applied to reproduce the complex double-peaked broad
emission lines and to reproduce the long-term variations of double-peaked emission 
lines, under the assumption that double-peaked broad emission lines come 
from disk-like BLRs locating onto central accretion disk. 

    For dbp emitters, because disk-like BLRs are locating onto accretion disks, 
one interesting result is that the disk-like BLRs can be partly obscured by 
dust torus in unified model (Antonucci 1993, Urry \& Padovani 1995) and/or 
by some dust clouds.  So, the interesting question is proposed whether 
broad emission lines of dbp emitters with partly obscured BLRs have 
similar properties as those of "normal" broad emission lines. In the 
following section, we develop our model and give some simple but interesting 
examples of broad emission lines with standard profiles from partly obscured 
disk-like BLRs of dbp emitters. In section 3, we try to find some candidates 
which have shifted standard broad emission lines, but actually are 
partly-obscured dbp emitters. Section 4 gives the discussion and conclusion.
The cosmological parameters $H_{0}=70{\rm km\cdot s}^{-1}{\rm Mpc}^{-1}$,
$\Omega_{\Lambda}=0.7$ and $\Omega_{m}=0.3$ have been adopted here.

\section{Theoretical Results From Partly Obscured BLRs Of dbp Emitters}

    The accretion disk model has been proposed and studied by many papers,
Circular disk model: Chen et al. 1989, Chen \& Halpern 1989,
Elliptical disk model: Eracleous et al. 1995, Warped disk model:
Bachev 1999, Hartnoll \& Blackman 2000, Circular disk with spiral arm
model: Hartnoll \& Blackman 2002, Karas et al. 2001.
Here, we select the elliptical accretion disk model (Eracleous et
al. 1995) rather than the other models, because the less parameters
of the model can better explain most of the aspects of double-peaked broad  
emission lines. The elliptical accretion disk model of Eracleous et al. (1995) 
has 8 free parameters. Five of them are applied to determine the geometrical
structure of the disk-like BLRs: the inner radius of disk-like BLRs $R_{in}$, 
the outer radius $R_{out}$, the eccentricity $e$, the inclination
angle $i$, and the original orientation angle of the BLRs in accretion 
disk $\phi_0$ (the angle between major axis of elliptical disk-like BLRs 
and direction of projected line-of-sight into accretion disk).
Another parameter $q$ is used to determine line emissivity as
$f_r\propto r^{-1\times q}$. Then two other parameters, local broadening velocity
dispersion $\sigma_{l}$ and amplitude factor $k$, are used to broaden
and strengthen outer broad emission lines. Here, we do not use the
parameter $k$. The normalized flux densities of broad H$\alpha$ from the
accretion disk model have a maximum value to 1.

   In order to discuss the effects of dust torus on observed broad emission 
lines from accretion disk model, some parameters of dust torus should be first 
determined. The size of torus has been determined by reverberation mapping 
technique applied to the flux variations in the near-infrared and optical 
bands for some nearby Seyfert 1 galaxies by Suganuma et al. (2006). The mean 
size of inner side of the torus for these nearby objects is about 
$\propto(\nu L_{\nu}(V))^{0.5}$, which is about three times larger than the 
size of BLRs of "normal" AGN as obtained by Kaspi et al. (2005). However, the 
opening angle of the torus cannot be determined by means of this unique 
parameter. If we accept that the size of BLRs of normal AGN should be less 
than the height of the inner side of dust torus, we can roughly determine 
the opening angle of the torus as about $40\ -\ 50$ degrees. 
Furthermore, the opening angle of torus can be estimated 
from the number ratio of type 1 to type 2 AGN (Zakamska et al. 2003). 
We can accept that opening angle of dust torus around 60 degree is a 
reasonable value. Thus if we adopt 60 degrees as the origin value of inclination 
angle in accretion disk model, $i=\pi/3$, which leads to the result that 
some part of disk-like BLRs should be seriously obscured by dust torus. 

\begin{figure*}
\centering\includegraphics[height = 18cm,width = 13cm]{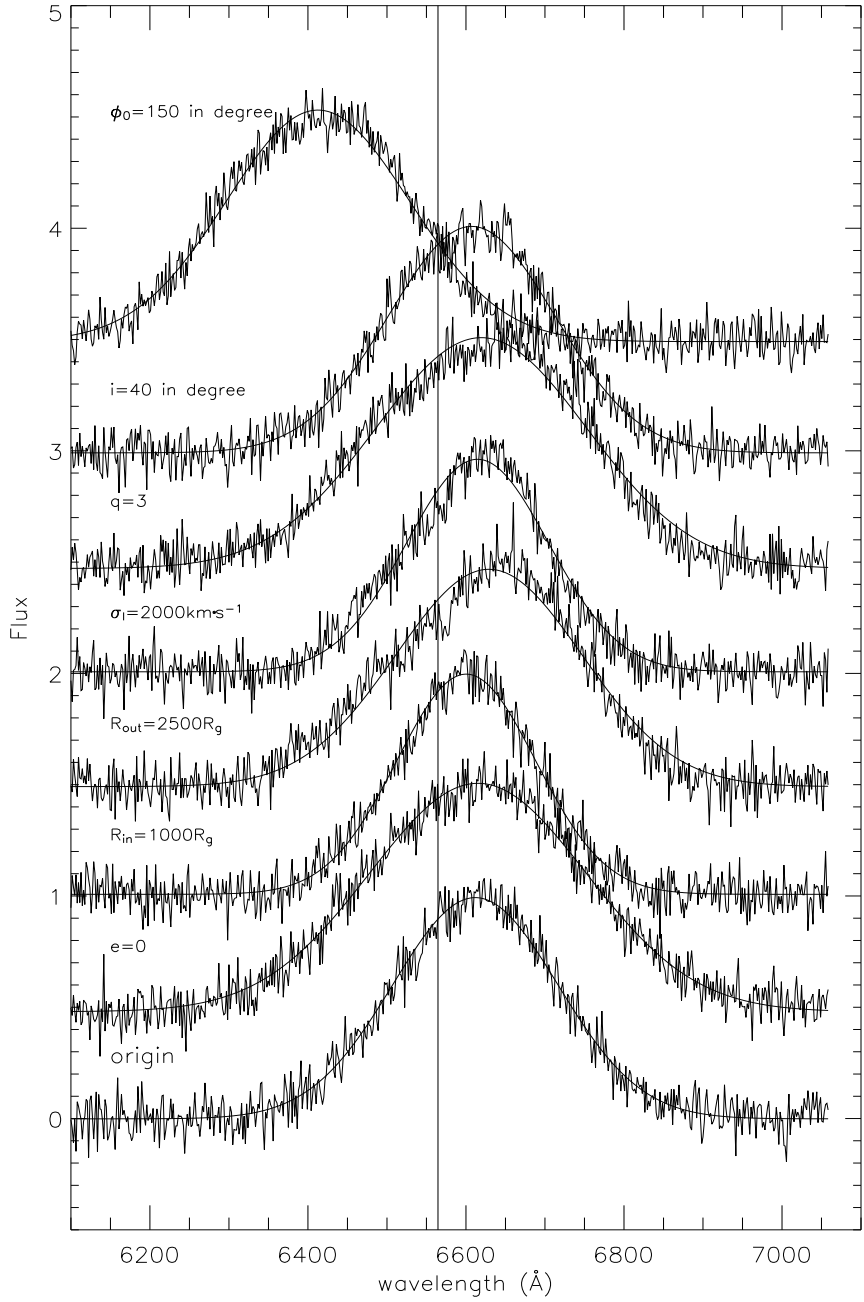}
\caption{ The created examples for broad H$\alpha$ by the partly 
obscured elliptical accretion
disk model. The thin line represents the created broad line by model.
Solid line represents the best fitted result by gaussian function.
The vertical line marks the position of the center wavelength of normal
broad H$\alpha$ in rest wavelength. For each created broad H$\alpha$,
the different input parameter is listed in the figure.
}
\label{exam}
\end{figure*}

  It is simple to chose the other model parameters to check output 
broad emission lines from partly obscured accretion disk model. The inner 
radius is about several hundreds of the gravitational radius,
$R_{in}\sim600 {\rm R_g}$. The outer radius is about several
thousands of the gravitational radius, $R_{out}\sim4000 {\rm R_g}$.
The local broadening velocity is several thousands of kilometers per second,
$\sigma_{l}\sim3000 {\rm km\cdot s^{-1}}$. The slope of the line
emissivity can be determined as $f_r\propto r^{-2}$. The eccentricity of
the elliptical disk is selected as $e = 0.6$ and the original orientation 
angle is $\phi_0\sim\pi/3$. The selected values for the model parameters 
(except local broadening velocity) above are common values for dbp emitters as 
shown in Eracleous \& Halpern (2003). Here, we select a bit larger local 
broadening velocity in order to get a clearly fine result as shown in 
Figure~\ref{exam}. As a simple example of the partly obscured accretion disk 
model, we consider the case that half of the H$\alpha$ emitting regions are 
seriously obscured, i.e., the integral range of orientation angle is 
not from $\phi_0$ to $\phi_0 + 2\times\pi$, but from $\phi_0 +\pi/2$ to 
$\phi_0 + 3\times\pi/2$.

Figure~\ref{exam} shows some output broad lines with some noise under the 
partly obscured accretion disk model. We can see that shifted standard 
gaussian broad H$\alpha$ emission lines by different input model parameters 
are produced. The random noise shown in Figure~\ref{exam} is created assuming 
that the maximum ratio of noise to flux density is less than 0.07, which is 
one standard value for SDSS spectra. We can find that different input model 
parameters lead to different line profiles with different shifted velocities, 
but can also be best fitted by standard gaussian function. Furthermore, the 
input model parameter of $\phi_0$ should determine if the observed broad 
H$\alpha$ is blue-shifted or red-shifted.

   Before the end of the section, it is interesting to check effects of 
the parameter of original orientation angle of partly obscured disk-like 
BLRs ($\phi_0$) on output broad line profile, because the original orientation 
angle is the only parameter in the elliptical accretion disk model which is 
varying with the precession of elliptical disk-like BLRs. The precession period 
should be proportional to $\frac{A\times(1-e^2)}{M_{BH}}$ ($A$ is the length 
of semi-major axis), for accretion disk around a Schwarzschild black hole,  
which indicates the precession for inner part of elliptical disk-like BLRs 
should be about several tens of years, and leads to apparent variations of 
the parameter of original orientation angle, if central BH masses are large 
enough and length of semi-major axis is small enough. Thus it is necessary 
to check the effects of original orientation angle on output broad emission lines 
under the partly obscured accretion disk model. Actually, in partly obscured 
accretion disk model, there are some free parameters which can not determined,  
such as actual area of obscured disk-like BLRs. Thus, there is so far no 
clear way to describe detailed effects of the original orientation angle. 
However, commonly we can simply show that the variation of original orientation 
angle should have little effects on the output model broad line profiles to some extent under the simple half partly obscured accretion disk model, although 
the half partly obscured accretion disk model above is oversimplified.

\begin{figure*}
\centering\includegraphics[height = 18cm,width = 13cm]{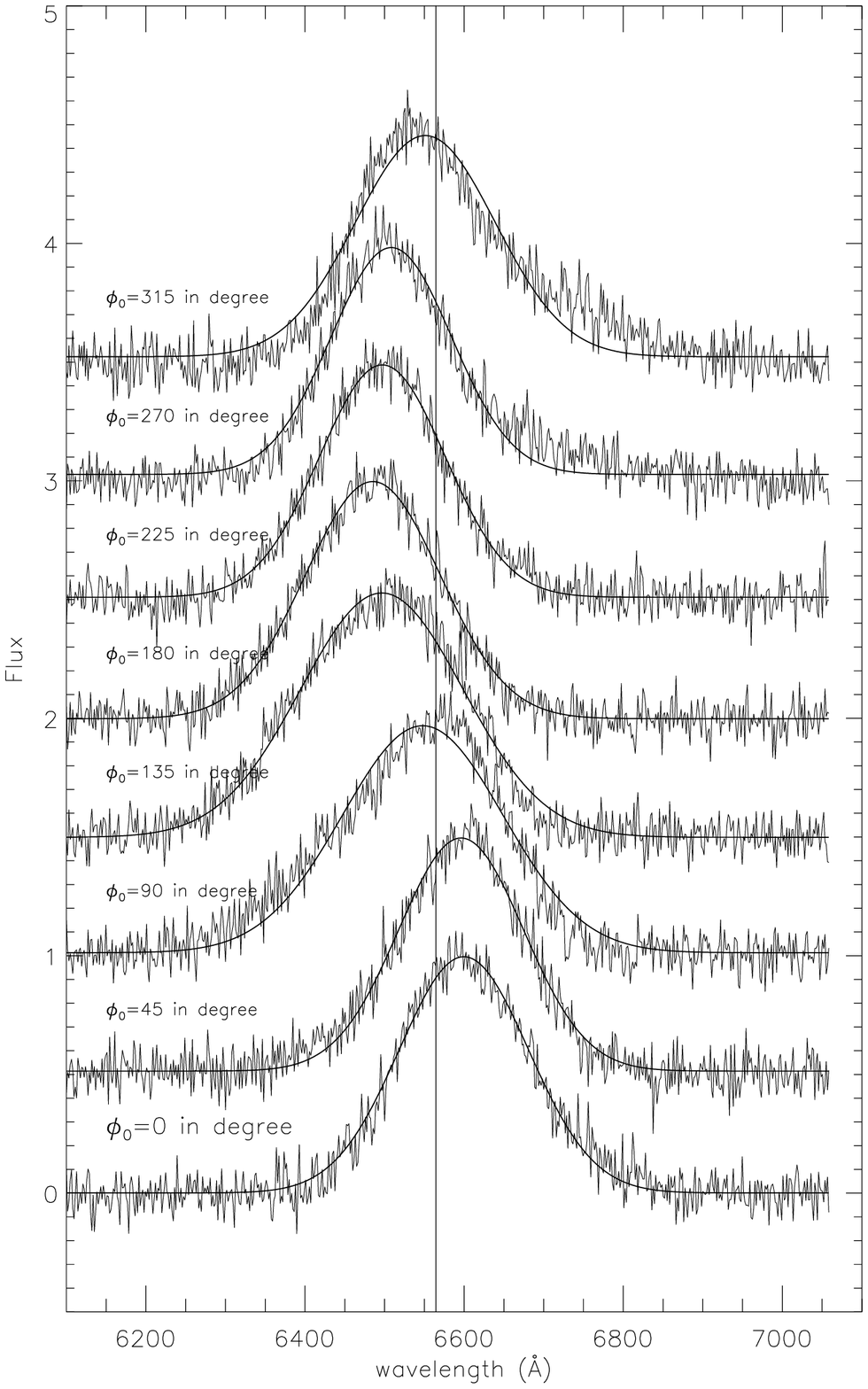}
\caption{ The created example for broad H$\alpha$ by the partly 
obscured elliptical accretion
disk model. The thin line represents the created broad line by model.
Solid line represents the best fitted result by gaussian function.
The vertical line marks the position of the center wavelength of normal
broad H$\alpha$ in rest wavelength. For each created broad H$\alpha$,
the different input parameter of orientation angle is listed in the figure, 
meanwhile the other parameters are held to constant to the values listed 
in text.}
\label{orian}
\end{figure*}

Figure \ref{orian} shows some examples of the output broad emission lines 
with different values of original orientation angle, meanwhile the other 
disk like parameters are held constant to the ones listed above.
There are eighty values of original orientation angle from 0 to $2\times\pi$ 
used to produce eighty output broad emission lines, and only 8 of them are 
shown in Figure \ref{orian}. It is clear that different orientation angles 
could lead to standard gaussian line profiles but with different shift 
velocities. Another interesting result about partly obscured accretion disk 
model is that the number of red-shifted model broad line is smaller than the 
number of blue-shifted model broad line, 25 red-shifted broad lines and 55 blue-shifted model broad lines, because of the elliptical disk-like BLRs with 
origin point at one focus point of the ellipse. Certainly, the number ratio 
should depends on the location of the central black hole, near focus 
point to the observer or far focus point to the observer. Detailed number ratio 
can be found in Section 3.2. The results shown in Figure \ref{orian} indicate 
that shifted standard gaussian broad emission line is not the case from some 
special selected original orientation angle.

  The results obtained above implies that some AGN (albeit a small number) 
with shifted broad emission lines with logarithmic profiles may well be
dbp emitters with partly obscured disk-like BLRs. There might be different cases
for partly obscured BLRs of dbp emitters, those in which a small part of
the BLRs are obscured and those where a large part the BLRs are obscured. 
If the central ring of BLRs is obscured, the expected result is that the peak
position should have a small shift velocity, because the rotation velocity
of the emission clouds in the outer rings is smaller than that in the
inner rings. Furthermore, if the obscured part is not seriously obscured 
by dust torus, the observed line profile should be seriously asymmetric.

\section{Observational Results From SDSS DR4}

   In this section, we discuss properties of some AGN selected from SDSS DR4 
with high qualified shifted standard gaussian broad H$\alpha$, in 
order to test our partly obscured accretion disk model.
We here focus on objects with higher signal-to-noise ratio than 40
at r band in the main quasars list of SDSS DR4 (Adelman-McCarthy et al.
2006), and with redshift less than 0.37 (in order to ensure the existence of 
complete broad H$\alpha$). Then, about 225 objects are firstly selected from 
SDSS according to criteria about signal-to-noise ratio and redshift.
We further check spectra of the selected objects and find that some of them 
have apparent stellar absorption features, such as absorption lines MgI$\lambda5175\AA$, CaII$\lambda3934, 3974\AA$ and 4000$\AA$ break.
Thus we make sure that before we measure line parameters, the stellar 
components in the observed spectrum are first subtracted.

    Before proceeding further, we first give some descriptions about our 
strict selection criteria. In this paper, we mainly focus on 
properties of shifted broad balmer emission lines. Thus how to determine 
whether one broad emission line is shifted is the first question we should 
find an answer. The common method is to calculate relative shifted value 
between central wavelengths of broad and narrow emission lines. It is 
obvious that asymmetry in broad emission lines should have serious effects 
on determination on central wavelength of broad emission lines. 
Thus, less asymmetry in high 
qualified emission lines can lead to better determination of shifted velocity. 
Besides the convenience to determine shifted velocity, shifted standard 
gaussian broad emission lines can provide more information about total 
isotropic radial motions of normal BLRs, if we accept that normal BLRs are 
locating into so-called ionization cone as described by Unified Model for 
AGN. In order to confirm observed broad H$\alpha$ has standard gaussian line 
profile, objects with high quality (S/N gt 40) are preferred. In order to 
confirm the broad emission lines have reliable shifted velocities, objects 
with apparent shifted velocities (larger than 180 $km/s$, the SDSS 
spectral resolution) relative to the center wavelength of narrow H$\alpha$ 
are selected. Thus number of objects in our final sample should 
be small, but the small number of objects still provide interesting and 
enough information about our final conclusion.

   An efficient method to subtract the stellar lights is the PCA
(Principle Component analysis) method described by Li et al.
(2005) and Hao et al. (2005), using the eigenspectra
from pure absorption galaxies from SDSS or the eigenspectra from stars
in STELIB (Le Borgne et al. 2003), because the method of Principle Component
Analysis (PCA) provides a better way to constrict more favorable
information from a series of spectra of stars or galaxies into several
eigenspectra. Here, we used the method from
Hao et al. (2005). The eigenspectra are calculated by KL (Karhunen-Loeve)
transformation for about 1500 pure absorption galaxies selected from SDSS
DR4. Then, the first eight eigenspectra and the spectra of an A star (which
is used to account for young stellar population) 
selected from STELIB (a library of stellar spectra at $R\sim2000$, Le Borgne 
et al. 2003) are used to fit the stellar properties of the observed spectra. 
After this,  rather than a power law, a three-order polynomial function is 
used to fit the featureless AGN continuum, because the study of composite 
spectra of AGN shows that the AGN continuum should be best fitted by two 
power laws with a break at $\sim5000\AA$ (Francis et al. 1991, Zheng et al. 1997,
Vanden Berk et al. 2001). After the last step, the featureless continuum and
the stellar components are obtained based on the Levenberg-Marquardt
least-squares minimization method applied for observed spectrum with emission 
lines masked.

   After the subtraction of stellar components and featureless continuum,
the line parameters can be measured. Here we focus on the region around 
H$\alpha$: broad and narrow components of H$\alpha$, [NII]$\lambda6548,6583\AA$ 
and [SII]$\lambda6716,6731\AA$. Then, we measure the line parameters 
applying gaussian function to each line component. In the procedure, the 
second moment of broad H$\alpha$ has a lower limit of 400${\rm km\cdot s^{-1}}$,  
the second moments of narrow emission lines have the same value in velocity space.
Once line parameters are measured, objects with shifted velocities larger than $180{\rm km/s}$ relative to narrow H$\alpha$ and/or [NII] doublet are selected. 
Then we carefully check the selected objects by eye to reject those  with 
double-peaked emission lines. We have about 20 objects, of which 
broad component of H$\alpha$ can be best fitted by one-gaussian
function. Then one criterion about $\chi^{2}$ is finally used to reject some 
objects. The parameter  $\chi^{2}$ is commonly used to determine
whether the model fit by one broad gaussian function is the best choice 
for broad emission lines limited by $0.5<\chi^{2}<2.5$, where $\chi^2$ is 
calculated by $\chi^2=\frac{\sum(\frac{y - y_{model}}{y_{err}})^2}{dof}$, 
where $dof$ is degree of freedom. We end up with  8 objects which have 
shifted standard gaussian broad H$\alpha$, seven with red-shifted broad 
H$\alpha$ and one with blue-shifted broad H$\alpha$. The best fitted results 
for emission lines around H$\alpha$ are shown in Figure \ref{ha}. In Table 1, 
we list the line parameters of the 8 objects.

   Before the end of the subsection, we should note that the broad component 
of H$\alpha$ in SDSS J1649 perhaps is not so secure, especially from the best 
fitted results shown in Figure \ref{ha}. However, through the measured line 
parameters of broad component of broad H$\alpha$ of SDSS J1649 (listed in 
Table 1), we also keep the object in our sample according to the criteria 
above.  

\subsection{BH Masses and Size of BLRs}

  The most reliable method to estimate BH masses is based on the stellar 
velocity dispersion of bulge of the host galaxy, M-sigma relation, first 
suggested by Ferrarese \& Merritt (2000) and Gebhardt et al. (2000), then 
confirmed by Tremaine et al. (2002) and Merritt \& Ferrarese (2001) etc.
\begin{equation}
M_{BH} = 10^{8.13\pm0.06}(\frac{\sigma}{200{\rm km\cdot s^{-1}}})^{4.02\pm0.32} {\rm M_{\odot}}
\end{equation}
which indicates strong correlation between BH masses and bulge masses
(H\"{a}ing \& Rix 2004, Marconi \& Hunt 2003, McLure \& Dunlop 2002,
Laor 2001, Kormendy 2001, Wandel 1999) etc. Furthermore, More 
recent results from a larger SDSS sample indicate there is no 
significant evolution of M-sigma relation by Shen et al. (2008). Besides, 
the results from observational results, Shankar et al. (2008) theoretically 
study the evolution of M-sigma relation, and found that from 
$z\sim0$ to $z\sim0.5$, there is no significant evolution of M-sigma 
relation, which is similar to the result found by Shen et al. (2008). 
Thus, although the eight objects in our sample have much different 
redshifts, we think the BH masses from M-sigma relation are still reliable.

   However, how to accurately measure stellar velocity dispersion is an
open question, because of known problems with the template mismatch.
A commonly used method is to select spectra of several kinds of
stars (commonly, G and K) as templates, and then broaden the templates
by the same velocity to fit stellar features, leaving the contributions
from different kinds of stars as free parameters
(Rix \& White 1992). However, more information about stars included
by the templates should lead to more accurate measurement of stellar
velocity dispersion. According to the above mentioned method to subtract
stellar components, we created a new template rather than several spectra
of G or K stars as templates. Thus, we apply the PCA method for all 255
spectra of different kinds of stars in STELIB. Selecting the first several
eigenspectra and a three-order polynomial function for the background as
templates, the value of stellar velocity dispersion can be measured by
the method of minimum $\chi^2$ method applied for the absorption features
around MgI$\lambda5175\AA$ within rest wavelength range from 5100$\AA$ to
5300$\AA$. The method to measure stellar velocity dispersion is similar to 
the method to subtract stellar component discussed above. In Figure \ref{abs}, 
we show the best fitted result for absorption features near MgI$\lambda5175\AA$ for
seven objects, except SDSS J1007+1246 because of the lack of
MgI$\lambda5175\AA$. For SDSS J1007+1246, the line width of narrow emission
lines is used as the substitute of stellar velocity dispersion
(Nelson \& Whittle 1995, Greene \& Ho 2005a),
$\sigma = 226.37 {\rm km\cdot s^{-1}}$.
Then BH masses of the eight objects are estimated by 
M-sigma relation and listed in Table 2.
Of course, we check the correlation between line width of narrow emission lines
and stellar velocity dispersion for the seven objects, and find the mean
value for the ratio of stellar velocity dispersion to line width of narrow emission lines (the second moment $\sigma$, for the standard gaussian line profiles, $FWHM = 2.35\times\sigma$) to be about $1.06\pm0.12$, which 
indicates that the measured stellar velocity dispersions for the seven 
objects is reasonably accurate. Due to the small number of objects, we do not 
show the correlation. The values of stellar velocity dispersion and line width 
of narrow emission lines are listed in Tables 1 and 2.

   The BH masses of the eight objects can be also estimated under the 
assumption of virialization method (Onken et al. 2004, Peterson et al. 2004, 
Kaspi et al. 2005, Bentz et al. 2006) and listed in Table 2,
\begin{equation}
M_{BH} = 2.36\times10^8(\frac{\sigma_B}{3000{\rm km\cdot s^{-1}}})^2(\frac{L_{5100\AA}}{10^{44}{\rm erg\cdot s^{-1}}})^{\sim0.5} {\rm M_{\odot}}
\end{equation}
Here, the more recent result about the correlation between 
size of BLRs and continuum luminosity is used (Bentz et al. 2006).
The AGN continuum luminosity is measured from the observed spectrum after the subtraction of stellar components and also listed in Table 2. As we 
have discussed in Zhang, Dultzin-Hacyan \& Wang (2007a), the equation above 
to estimate virial BH masses of AGN is not accurate for some dwarf AGN with 
much lower dimensionless accretion rate $\dot{m_{H\alpha}} = \frac{L_{H\alpha}}{L_{Edd}}<10^{-5.5}$ (where $L_{H\alpha}$ includes both the broad and 
narrow components of H$\alpha$), because of the inaccuracy of the empirical 
elation between continuum luminosity and size of the BLRs. Thus, the virial 
BH mass of the object SDSS J1649+3613, with $\dot{m_{H\alpha}}\sim10^{-6}$ should 
be smaller than the determined value because  $R_{BLRs}$ estimated from 
continuum luminosity should be smaller than the true size (see also 
Wang \& Zhang 2003).

   If the shifted standard gaussian broad H$\alpha$ comes from the partly 
obscured accretion disk, we should expect that the virial BH masses should 
be smaller than the more reliable BH masses estimated through stellar velocity dispersions, because of the smaller observed line width and smaller size of BLRs 
from smaller observed continuum luminosity than intrinsic ones from partly 
obscured accretion disk model. Figure \ref{mass2} shows the comparison of the 
two kinds of BH masses of the eight objects. From the figure, we can see that 
BH masses estimated through M-sigma relation are systemically larger than virial 
BH masses, $<\frac{M_{BH}(\sigma)}{M_{BH}(virial)}>\sim68$ for all the eight 
object (If the object SDSS J1649 is rejected, the mean ratio of the two kinds of
BH masses is 16). The result strongly indicates that the partly obscured 
accretion disk model is preferred to the objects in our sample, especially 
SDSS J1044, SDSS J1457 and SDSS J1715. 

   It is obvious that we do not consider effects from internal reddening on 
results above. Commonly, effects of internal reddening can be
corrected through balmer decrements from broad balmer emission lines, 
especially flux ratio of broad H$\alpha$ to broad H$\beta$, 
assuming internal balmer decrement for broad H$\alpha$ to broad H$\beta$ 
is about 3.1. The flux ratios of broad H$\alpha$ to broad H$\beta$ are 
listed in Table 2 for the eight objects. According to observed flux ratios 
of H$\alpha$ to H$\beta$, internal luminosity of broad and H$\alpha$ 
are determined, after the correction of internal reddening effects. Then 
according to the correlation between continuum luminosity and line luminosity 
for a sample of quasars (Greene \& Ho 2005b), internal AGN continuum luminosity 
is estimated for each object in our sample. In other words, after the correction 
of internal reddening effects, we ensure that flux ratio of H$\alpha$ to H$\beta$ 
is 3.1, and the correlation luminosity of H$\alpha$ and continuum luminosity 
obeys the one found in Greene \& Ho (2005b). According to the AGN continuum luminosity after the correction of internal reddening effects, the virial BH 
masses are re-measured by Equation (2), and re-shown in Figure \ref{mass2n}. 
We can clearly see that the effects of internal reddening cannot 
change the result: BH masses estimated through stellar velocity are systemically larger than virial BH masses. 

   If we assume that the broad H$\alpha$ of each object in our sample is
emitted from "normal" BLRs, its distance to the BH (its size) can be estimated
from the continuum luminosity after the correction of internal reddening 
(Bentz et al. 2006), $R_{BLRs}\propto L_{5100\AA}^{\sim0.5}$,
except for SDSS J1649 which has much smaller
dimensionless accretion rate $\dot{m_{H\alpha}}<-5.5$
(Zhang, Dultzin-Hacyan \& Wang 2007a).
The estimated size of BLRs of each object is listed in Table 2. 
If the shifted broad H$\alpha$ are interpreted by emitting regions dominated 
by radial motions, it is clear that radial motions should lead to change of 
size of BLRs, and then lead to change of line width. There are 7 out of 8 
objects show red-shifted broad profiles. The red-shifted velocity indicates 
the direction of the radial flows points to the central black hole. 
The inferred "infalling times" for such emission clouds were estimated and
are listed in Table 2. The implication is the following: If the shift of broad H$\alpha$ is due to infalling emission clouds, after several years the broad 
line emission clouds would be accreted into the central accretion disk or into 
the central black hole, and thus the observed broad H$\alpha$ should have 
complex line profiles rather standard gaussian profiles.

    Finally, we want to point out that the smaller line width of
broad H$\alpha$ for objects in our sample is about 30$\AA$ (second moment). 
It is necessary to check whether the partly obscured disk model can 
reproduce the outward shifted standard gaussian broad H$\alpha$ with small 
line width. If we select the following model parameters:
$R_{in}\sim6286 {\rm R_g}$, $R_{out}\sim45000 {\rm R_g}$,
$i=60\degr$, $f_r\propto r^{-1.5}$, $e = 0.8$, $\sigma_l\sim1550 {\rm km\cdot s^{-1}}$ and $\phi_0\sim55\degr$, the outward gaussian broad H$\alpha$ line 
profile of SDSS J1715+5935 can be nearly reproduced with a second moment of 38.28$\AA$ and with center wavelength at 6577.92$\AA$.  For this we have to 
assume that half of the disk-like BLRs in accretion disk are obscured. 
This is, however, a very particular assumption to explain a particular case.
By changing of the input model parameters, outward broad line with different
line width and different shifted velocity can be obtained. However, we should
notice that this is a simple  model, where we assume that half of the broad
dbp clouds  are heavily obscured by the dust torus. Probably the actual case
is much complex. Thus here we do not give specific model parameters for the
objects in our sample, our main objective is to show that partly obscured 
dbp BLRs can lead to shifted broad lines with normal form.

\subsection{Number Ratio of Objects with Red-Shifted Velocities to 
Objects with Blue-Shifted Velocities}

   In this subsection, we will discuss the number ratio of 
objects with red-shifted observed broad standard gaussian H$\alpha$ to 
objects with blue-shifted observed broad standard gaussian H$\alpha$, 
$N_{rb}$,  under the partly obscured accretion disk model, which will provide more evidence for the model. 

  As what we have done in section 2, we can find that under the 
partly obscured accretion disk model, the value of $N_{rb}$ is about 
$\frac{25}{55}$. It is clear that the value should depends on the 
location of central black hole in the accretion disk model, 
if the radius of one particle is described by $r=\frac{r_{\star}(1+e)}{1+e\times\cos(\phi-\phi_0)}$ (where $r_{\star}$ 
and $e$ are the pericenter distance and eccentricity of one 
elliptical ring), rather than by $r=\frac{r_{\star}(1+e)}{1-e\times\cos(\phi-\phi_0)}$, (in other words, 
the new orientation angle $\phi^{\star}$ is described 
as $\phi^{\star}=\phi+\pi$), the 
value of $N_{rb}$ should be $\frac{55}{25}$. Especially, if the 
eccentricity is around zero, i.e., circular disk-like BLRs, it is 
obvious that the number ratio of $N_{rb}$ should be 1.
Thus the number ratio from the partly obscured accretion disk model 
is around 1.

   However, there is only one object with blue shifted standard 
broad gaussian H$\alpha$. Perhaps the weird number ratio is due to 
the strict selection criteria. Figure \ref{nrat} shows the 
distribution of shifted velocities ($w_{0}(H\alpha_B) - w_{0}(H\alpha_N)$) of all broad line AGN with reliable standard gaussian broad H$\alpha$. 
The value of $N_{rb}$ is about 1.02 for the 1672 selected broad line 
AGN. Simply procedure to select 
the objects as follows. The emission lines around H$\alpha$ are 
fitted twice. Firstly, each gaussian function is applied to each 
emission line, broad and narrow emission lines, for all the 8668 
QSOs in SDSS DR6. Then the parameter of $\chi^2_{1}$ 
is measured. Secondly, double broad gaussian 
functions are applied to broad H$\alpha$, with measured parameter 
$\chi^2_{2}$. Then only the objects with similar values of $\chi^2_1<2$ 
and $\chi^2_2<2$, $\chi^2_{1} - \chi^2_{2}\le0.1$, are selected. The similar values of $\chi^2_1$ and $\chi^2_2$ mean that 
the double gaussian functions for broad H$\alpha$ is meaningless. 
Thus the weird number ratio for the objects listed in Table 1 and in 
Table 2 is due to the strict selection criteria, not a intrinsic 
phenomenon.

Furthermore, we check the correlation between virial BH masses 
estimated from equation (2) and BH masses from M-sigma relation in 
equation (1) for the 216 objects with reliable measured stellar 
velocities through absorption features around MgI$\lambda5175\AA$, 
and with shifted velocities larger than 180 ${\rm km/s}$. 
Figure \ref{mass2n} shows the results. We can find what we have expected 
from partly obscured accretion disk model: 
the BH masses from M-sigma relation are one magnitude higher than the 
virial BH masses for the objects with shifted velocities larger than 
180${\rm km/}s$,  $<\frac{M_{BH}(\sigma)}{M_{BH}(virial)}>\sim17$. 

To sum up, if the selection criteria are not so strict, the number 
ratio of $N_{rb}$ should be around 1, as shown in Figure \ref{nrat}. 
However, for low quality SDSS spectra with smaller S/N, 
it is difficult to confirm the observed broad H$\alpha$ have 
standard gaussian profiles, because of the strong noise in the spectra. 
Certainly, it is reasonable to obtain some statistical results through 
the large sample of objects with low quality, as shown in 
Figure \ref{mass2n}. Certainly, the radial motions in BLRs for 
shifted broad emission lines are not rejected, because some 
objects have coincident virial BH masses and BH masses from 
M-sigma relation. However we provide an optional model to explain 
the appearance of shifted broad emission lines, especially 
for those objects with underestimated virial BH masses. Certainly, 
there are some observed evidence which can be used to determine 
which model, the partly obscured accretion disk model or radial motions 
in BLRs model, should be preferred for AGN with under estimated virial BH 
masses, as discussed in the following section.

\section{Discussions}

   In this paper we present a new and interesting model to explain the 
appearance of shifted broad emission lines, besides the radial motions of broad emission line clouds in BLRs. The scenario considers the partly obscuration of 
disk-like BLRs in the accretion disk of dbp emitters. From the shift velocity of broad line, we cannot
determine velocity field of the broad line clouds in BLRs. We cannot firmly discard radial motions into the Black Hole or away from it.  But we can predict that if the shifted standard gaussian broad lines are emitted from a  
"normal" BLRs in an ionization cone, we expect the line width to change with
time passage (BLRs near the central Black Hole produce broader emission
lines), but the line profile should keep the its form. As shown in Table 2,
the infalling times for several objects are only around ten years, thus it 
should be possible to determine whether radial motions are dominant in these objects by observational spectroscopic monitoring.  On the other hand if 
our model of partly obscured BLRs of dbp emitters is the correct one, we 
expect that the shapes of the line profiles for some special cases will change 
with time passage, especially the peak intensity, if the precession of 
elliptical disk-like BLRs is not so longer than several hundreds of years. 
Because we only have single epoch spectra, it is difficult to determine the
model parameters for the objects in our sample. Thus here we do not give the
precession period of BLRs into accretion disk for dbp emitters, and can not give
a clear time prediction for observing the change in line profiles. Thus the
long-period monitoring of these objects is an interesting observational 
project to confirm our model. 

   Furthermore, besides the shifted standard broad gaussian emission lines, 
it is interesting to discuss the objects with asymmetric broad emission lines. 
Commonly, the asymmetry is considered as the effects of radial motions 
in part of normal BLRs. However, 
the asymmetric non-virialized components in broad emission lines should lead 
to overestimated virial BH masses. Thus it is clear that there are 
obviously different characters of virial BH masses estimated from the 
radial motions in BLRs model and from the partly obscured accretion disk model. 
The radial motions in BLRs always indicates overestimated virial BH masses, 
however, partly obscured accretion disk model leads to underestimated virial 
BH masses.

\vspace{6mm}
   Finally, we can resume our conclusions as follows: We present a model to
explain the appearance of shifted Balmer broad emission lines which does not need to involve radial motions dominating the emission line clouds. We show that partly obscured BLRs of dbp emitters can also produce shifted
standard Gaussian broad emission lines. Then we select eight high quality 
objects (with S/N at r band larger than 40) with shifted broad 
H$\alpha$ (the shifted velocity larger than 180 ${\rm km\cdot s^{-1}}$) 
from SDSS DR4 (seven of them have observable stellar absorption features). Reliable BH masses determined using stellar velocity dispersions 
result systemically larger than virial BH masses estimated by line width 
of broad H$\alpha$ and continuum luminosity. We further estimate the size of 
BLRs for the objects from continuum luminosity and show that internal 
reddening has no important influence on our final results.
Finally, we make predictions about the variation of shifted broad H$\alpha$ with  time passage for the two models: Broad H$\alpha$ from "normal" radially moving clouds or broad H$\alpha$ from dbp BLRs in a partly obscured accretion disk.

\section*{Acknowledgements}
ZXG gratefully acknowledges the postdoctoral scholarships offered by
Max-Planck Institute f\"ur Astrophysik. D. D acknowledges
support from grant IN100507 from DGAPA, UNAM.
This paper has made use of the data from the SDSS projects.
Funding for the creation and the distribution of the SDSS Archive
has been provided by the Alfred P. Sloan Foundation, the
Participating Institutions, the National Aeronautics and
Space Administration, the National Science Foundation,
the U.S. Department of Energy, the Japanese Monbukagakusho, and the
Max Planck Society. The SDSS is managed by the Astrophysical Research
Consortium (ARC) for the Participating Institutions. The Participating
Institutions are The University of Chicago, Fermilab, the
Institute for Advanced Study, the Japan Participation Group,
The Johns Hopkins University, Los Alamos National Laboratory,
the Max-Planck-Institute for Astronomy (MPIA),
the Max-Planck-Institute for Astrophysics (MPA), New
Mexico State University, Princeton University, the United
States Naval Observatory, and the University of Washington.

\begin{table*}
\centering
\begin{minipage}{160mm}
\caption{Parameters of emission lines}
\begin{tabular}{llllllllll}
\hline
id & Name & z  & w$_{0, N}$ ($\AA$) & $\sigma_N$ ($\AA$) & flux$_N$ &
w$_{0, B}$ ($\AA$) & $\sigma_B$ ($\AA$) & flux$_B$ & $\chi^2$ \\
\hline
0 & J093943.74+560230.4 &  0.116 &  6566.70$\pm$0.23 &  3.94$\pm$0.31 &  103.059$\pm$10.93 &  6557.62$\pm$0.18 &  44.50$\pm$0.19 &  7321.52$\pm$34.17 &  1.79 \\
1 & J100726.10+124856.2 &  0.240 &  6562.47$\pm$0.17 &  4.95$\pm$0.21 &  827.151$\pm$31.22 &  6576.65$\pm$0.47 &  65.41$\pm$0.66 &  14279.4$\pm$145.2 &  0.83 \\
2 & J102044.43+013048.4 &  0.096 &  6565.87$\pm$0.02 &  1.77$\pm$0.02 &  310.504$\pm$4.708 &  6574.76$\pm$0.54 &  49.74$\pm$0.64 &  1728.50$\pm$23.29 &  1.12 \\
3 & J104451.73+063548.6 &  0.027 &  6565.12$\pm$0.03 &  3.86$\pm$0.04 &  1522.16$\pm$34.03 &  6570.33$\pm$0.39 &  18.49$\pm$0.28 &  2251.47$\pm$68.52 &  1.99 \\
4 & J145706.79+494008.4 &  0.013 &  6565.75$\pm$0.04 &  2.41$\pm$0.04 &  1902.73$\pm$40.37 &  6570.76$\pm$0.21 &  21.15$\pm$0.19 &  6961.25$\pm$79.99 &  1.12 \\
5 & J164909.58+361325.8 &  0.030 &  6566.17$\pm$0.09 &  6.31$\pm$0.12 &  1626.77$\pm$50.02 &  6571.28$\pm$1.04 &  26.12$\pm$1.68 &  1026.26$\pm$105.3 &  1.46 \\
6 & J171322.58+325628.0 &  0.101 &  6565.42$\pm$0.07 &  4.21$\pm$0.09 &  461.218$\pm$9.292 &  6571.57$\pm$0.33 &  43.90$\pm$0.45 &  2594.12$\pm$24.25 &  1.01 \\
7 & J171550.49+593548.7 &  0.065 &  6565.48$\pm$0.07 &  3.68$\pm$0.08 &  202.362$\pm$5.110 &  6577.68$\pm$0.24 &  37.16$\pm$0.28 &  1879.38$\pm$14.48 &  2.18 \\
\hline
\end{tabular}
\\
Notes:--\\
First column is the name of each object in the format of
'Jhhmmss.ss$\pm$ddmmss.s', second column is the redshift. Third to fifth
columns list the line parameters of narrow H$\alpha$, center wavelength
in unit of $\AA$, line width in unit of $\AA$ and flux in unit of
$10^{-17} {\rm erg\cdot s^{-1}\cdot cm^{-2}}$ of narrow H$\alpha$.
Then the line parameters of broad H$\alpha$ are listed in
next three columns, center wavelength in unit of $\AA$, line width in
unit of $\AA$ and flux in unit of
$10^{-17} {\rm erg\cdot s^{-1}\cdot cm^{-2}}$ of broad H$\alpha$.
The value of $\chi^2$ is listed in the last column.
\end{minipage}
\end{table*}

\begin{table*}
\centering
\begin{minipage}{160mm}
\caption{Parameters of emission lines}
\begin{tabular}{llllllllll}
\hline
id & name & $\sigma$ & $\dot{m_{H\alpha}}$ & BD & $L_{5100\AA}$ & $M_{BH}(\sigma)$ &
$M_{BH}(V)$ & $R_{BLRs}$ & $t_{in}$ \\
\hline
0 &  J093943.74+560230.4 &  165.52$\pm$26.62 &  -3.5 &  3.67 &  43.85 &  7.79 &  7.96 &  20.24 &   \\
1 &  J100726.10+124856.2 &  226.37$\pm$7.17 &  -3.0 &  5.14 &  44.94 &  8.34 &  8.89 &  80.49 &  102.1 \\
2 &  J102044.43+013048.4 &  137.64$\pm$15.36 &  -3.9 &  3.89 &  43.23 &  7.47 &  7.71 &  9.23 &  18.6 \\
3 &  J104451.73+063548.6 &  148.48$\pm$11.53 &  -4.9 &  15.13 &  43.29 &  7.61 &  6.88 &  9.96 &  34.3 \\
4 &  J145706.79+494008.4 &  140.43$\pm$12.26 &  -5.1 &  4.14 &  42.29 &  7.51 &  6.45 &  2.81 &  10.1 \\
5 &  J164909.58+361325.8 &  264.23$\pm$15.46 &  -5.9 &  12.68 &  42.98 &  8.61 &  6.82$^{\star}$ &  6.72 &  23.7 \\
6 &  J171322.58+325628.0 &  171.56$\pm$20.50 &  -4.1 &  3.64 &  43.38 &  7.86 &  7.68 &  11.15 &  32.7 \\
7 &  J171550.49+593548.7 &  151.30$\pm$16.15 &  -4.4 &  4.31 &  43.01 &  7.64 &  7.33 &  6.98 &  10.3 \\
\hline
\end{tabular}
\\
Notes:--\\
The id number and name for each object is listed in the first and second
column. Third  column gives
the value of stellar velocity dispersion in unit of ${\rm km\cdot s^{-1}}$.
Forth column is the dimensionless accretion rate by luminosity of H$\alpha$ before the correction of internal reddening for BLRs $\dot{m_{H\alpha}}$. The flux ratio
of broad H$\alpha$ to broad H$\beta$ is listed in Column V. Then the internal
logarithmic continuum luminosity at 5100$\AA$ in unit of
${\rm erg\cdot s^{-1}}$ is shown in the next column, after
consideration of the effects of internal reddening. Column VII - VIII are the two kinds of logarithmic BH masses in unit of ${\rm M_{\odot}}$, $M_{BH}(\sigma)$
is estimated from stellar velocity dispersion, $M_{BH}(V)$ is the virial BH
masses through line width of broad H$\alpha$ and continuum luminosity listed
in Column VI. For SDSS J1649, the listed virial BH mass in the table is 
the low-limitted one.
The last two columns are the size of BLRs in unit of light-days estimated by the continuum luminosity listed in Column VI and the infalling time
in unit of years.
\end{minipage}
\end{table*}

\begin{figure*}
\centering\includegraphics[height = 6cm,width = 8cm]{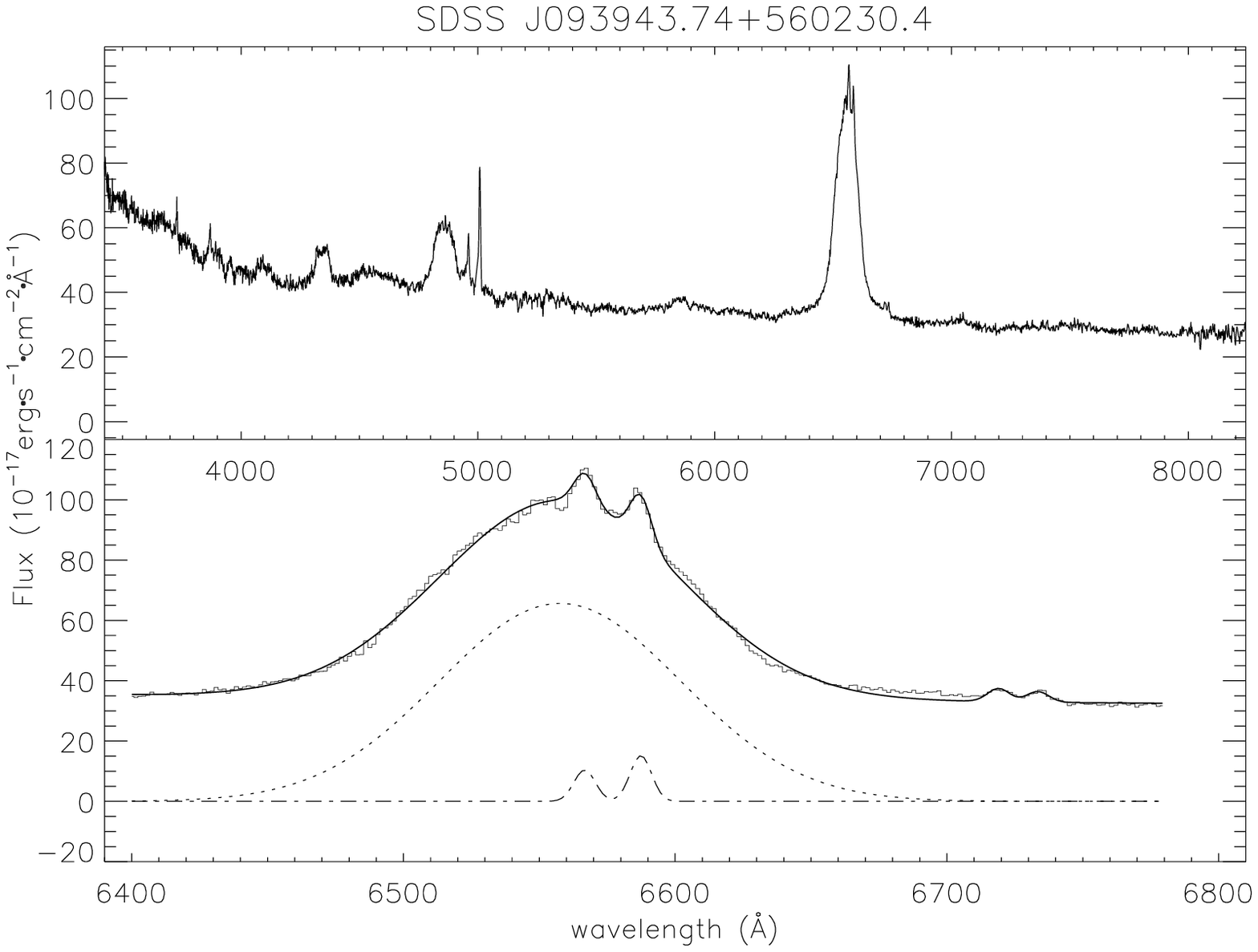}
\centering\includegraphics[height = 6cm,width = 8cm]{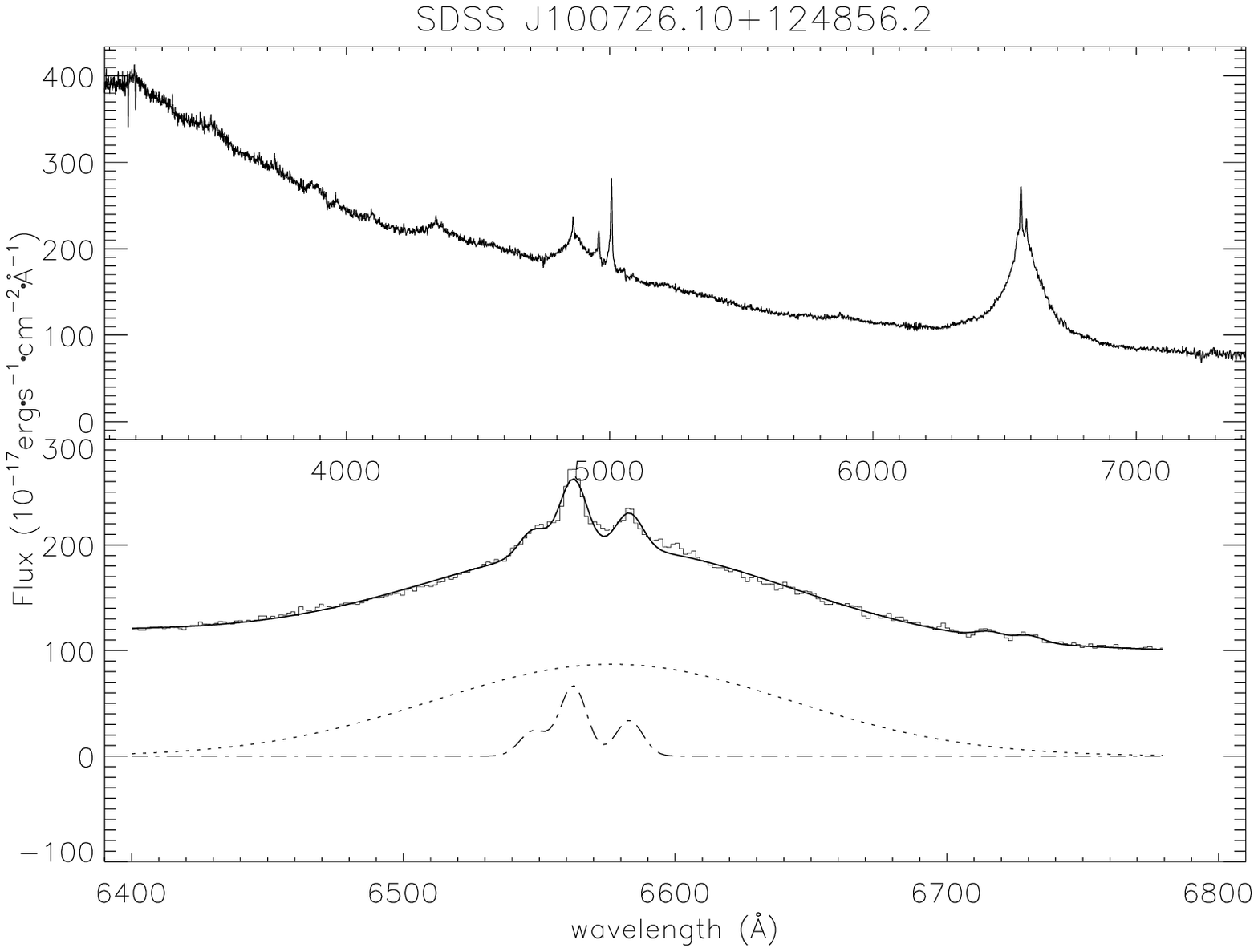}
\centering\includegraphics[height = 6cm,width = 8cm]{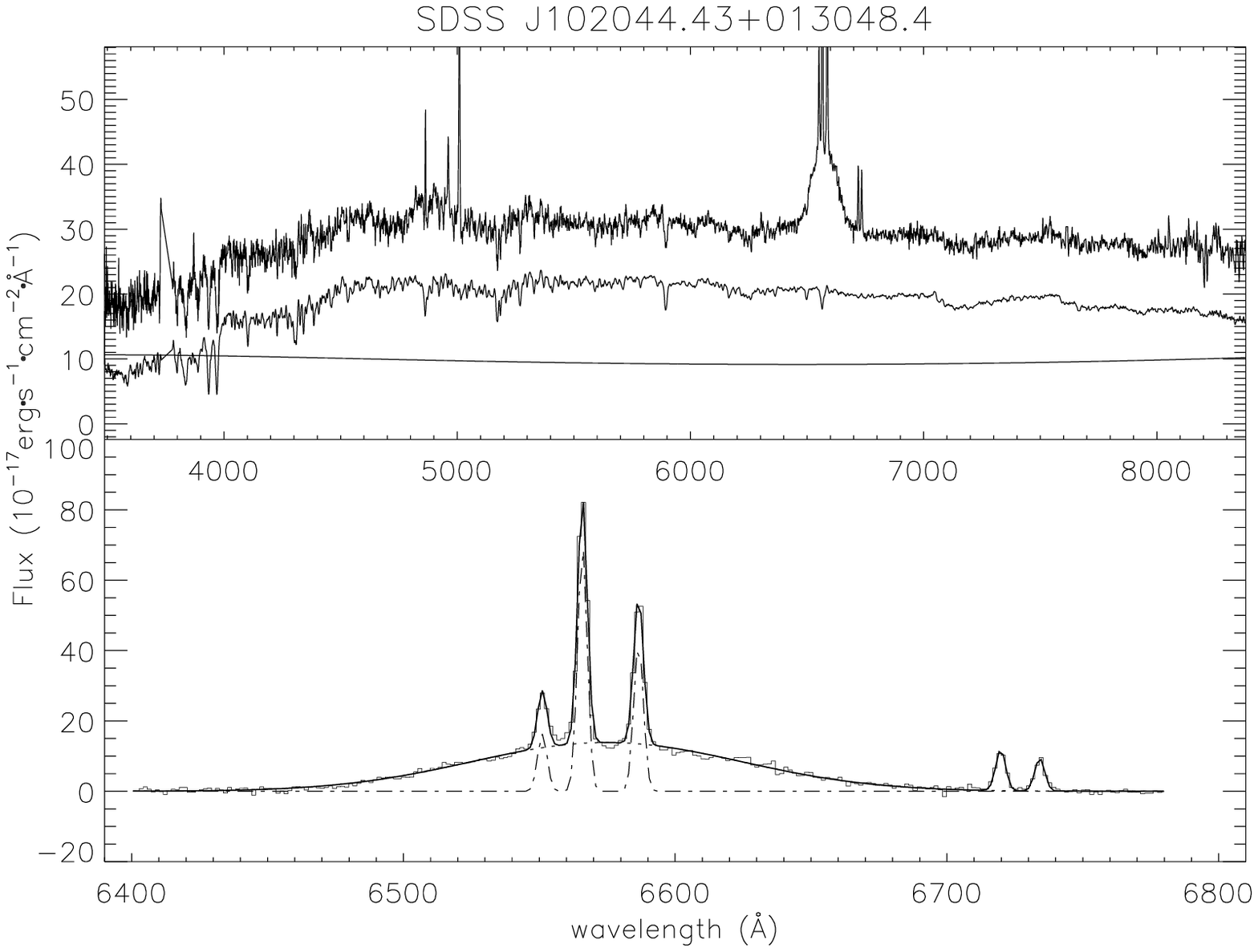}
\centering\includegraphics[height = 6cm,width = 8cm]{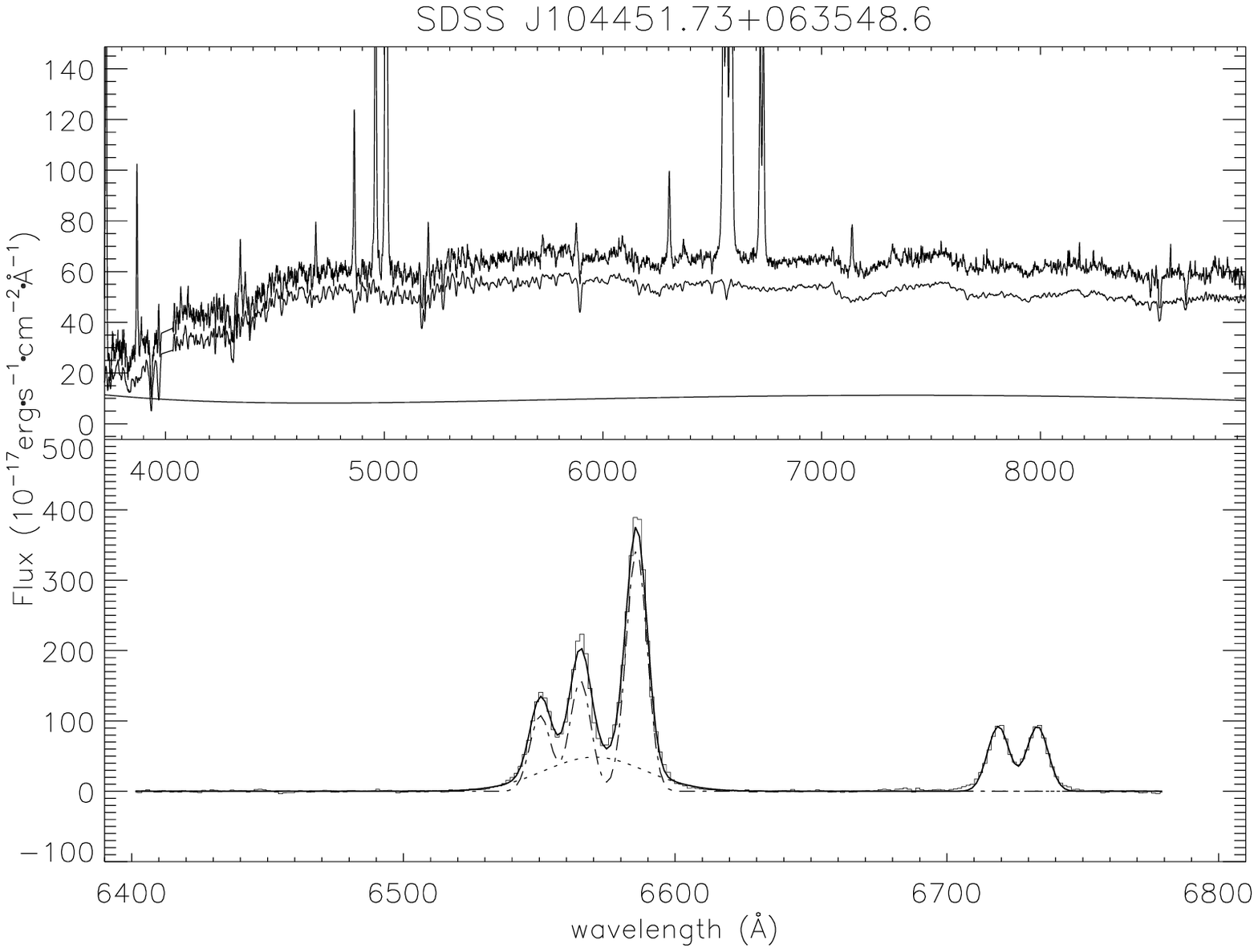}
\centering\includegraphics[height = 6cm,width = 8cm]{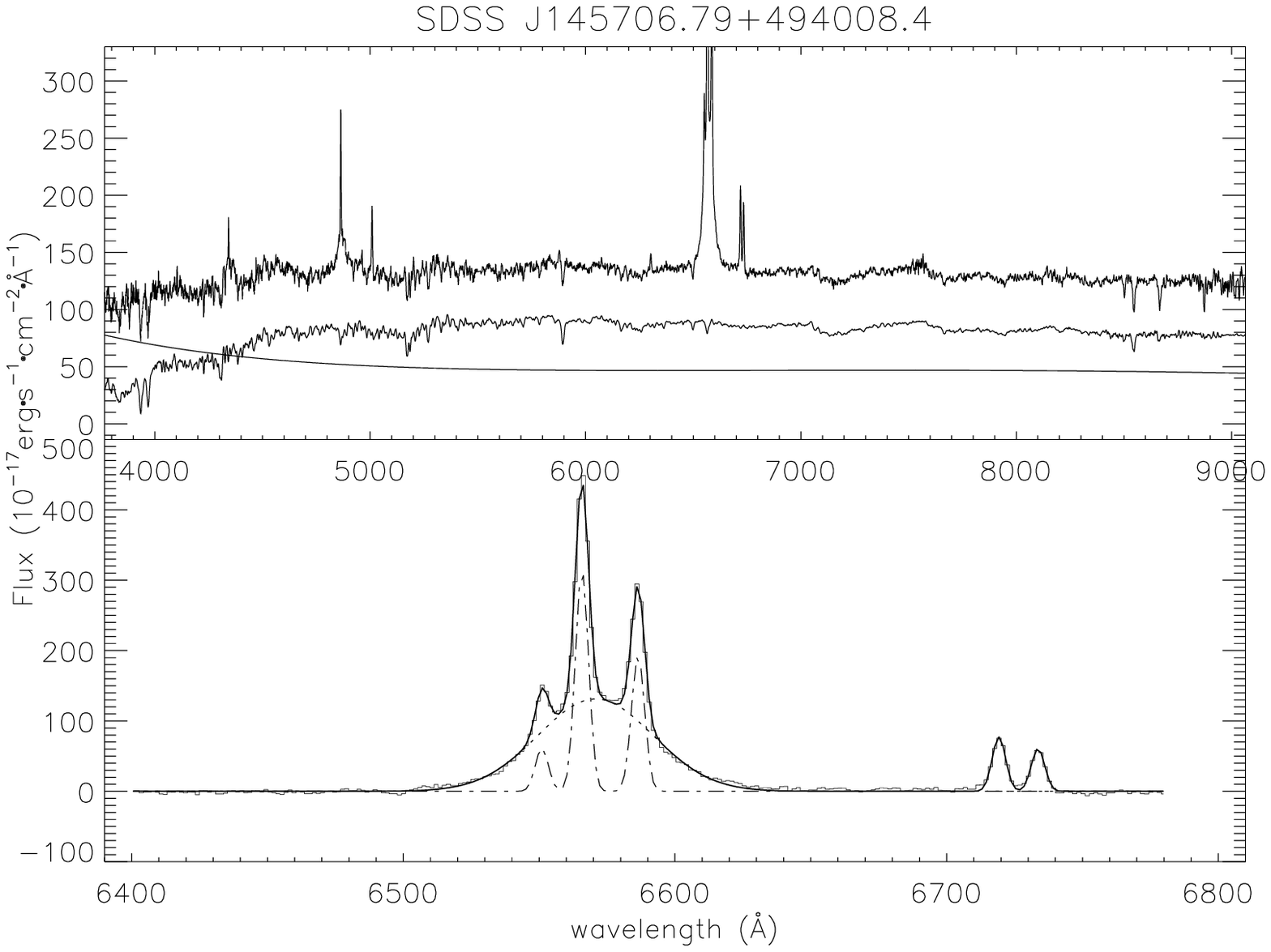}
\centering\includegraphics[height = 6cm,width = 8cm]{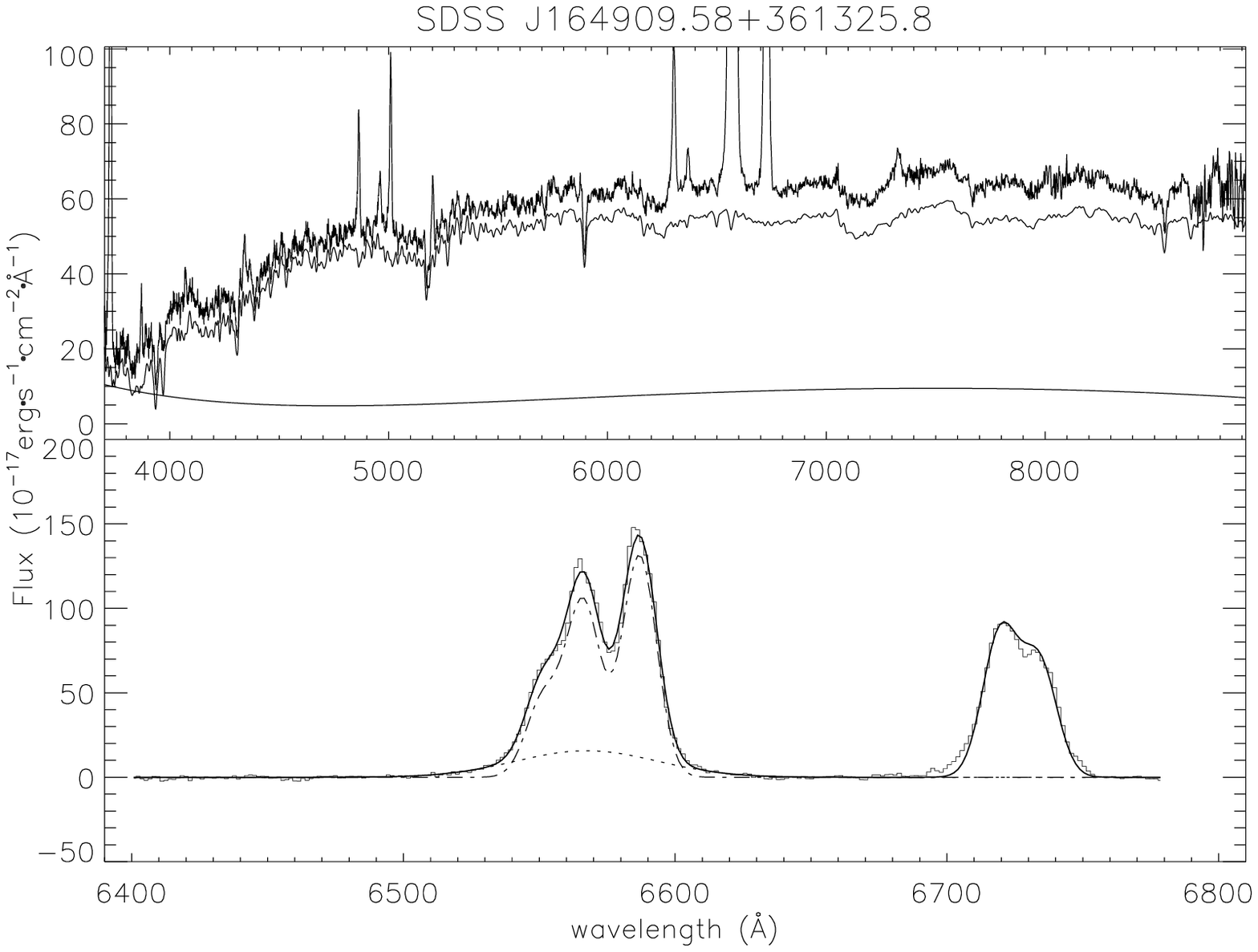}
\caption{The best fitted results for emission lines around H$\alpha$ are
shown in bottom panel for each object. In the panel, thin solid line
represents the observed spectrum, thick solid is for the best fitted
results, dotted line is for the standard gaussian broad H$\alpha$,
dot-dashed line represents the narrow components of H$\alpha$ and
[NII]$\lambda6548,6583\AA$.
In top panel for each object, the observed spectrum is shown. If there
is apparent stellar features, the features and featureless continuum
emission are also shown in the panel. If there is no stellar features,
only the observed spectrum is shown in the panel. The best fitted
}
\label{ha}
\end{figure*}

\setcounter{figure}{2}
\begin{figure*}
\centering\includegraphics[height = 6cm,width = 8cm]{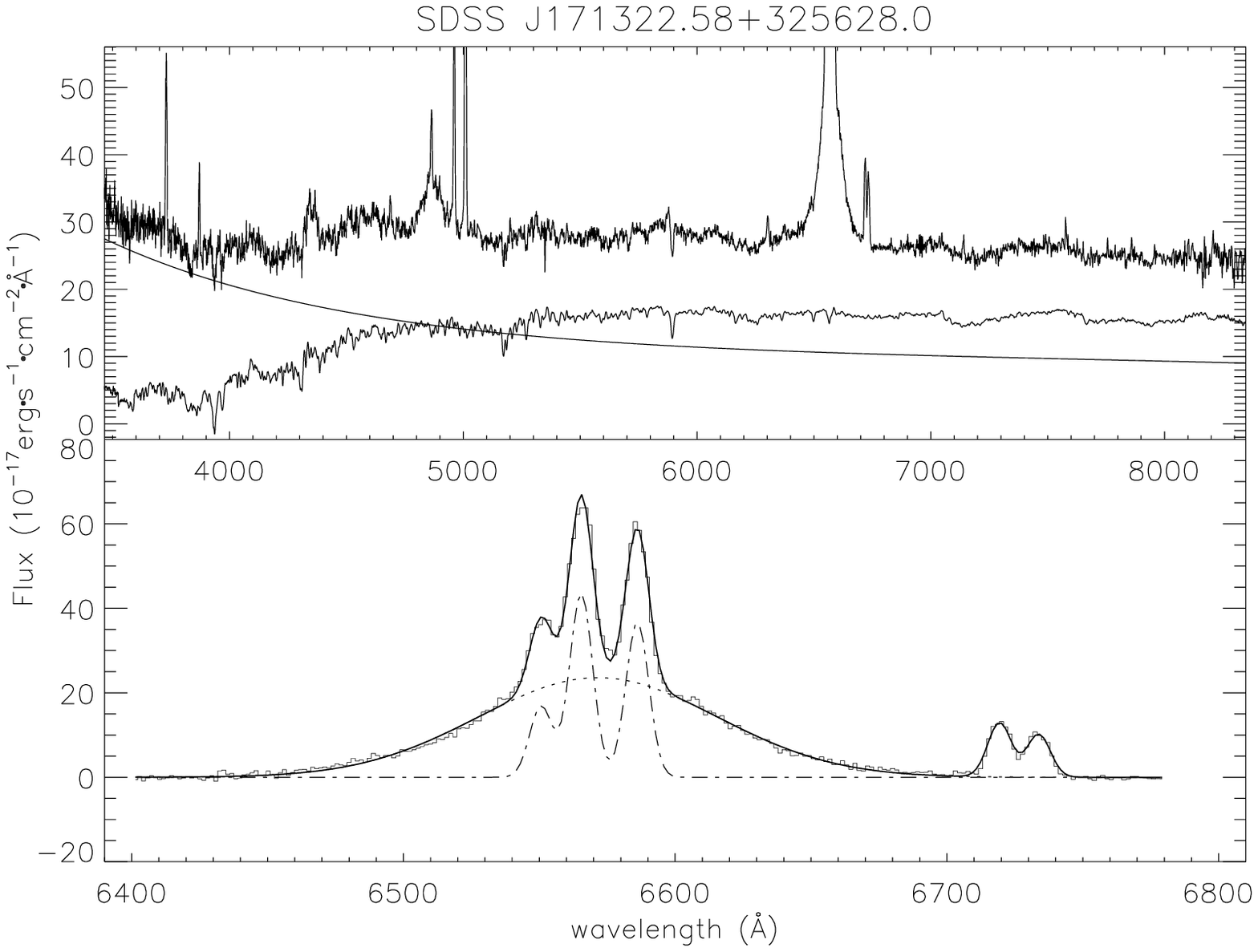}
\centering\includegraphics[height = 6cm,width = 8cm]{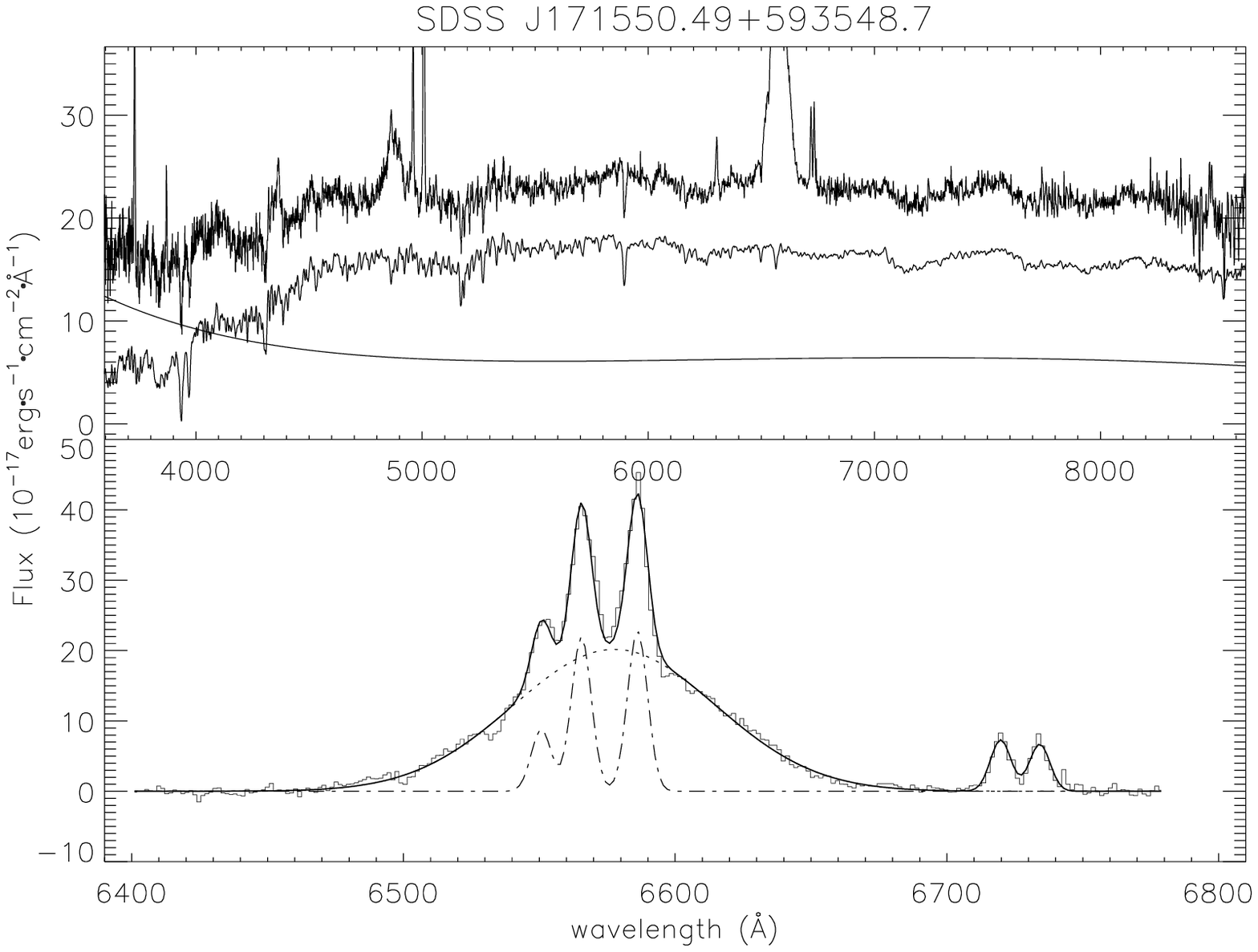}
\caption{ -- Continued.
}
\end{figure*}

\begin{figure*}
\centering\includegraphics[height = 5cm,width = 8cm]{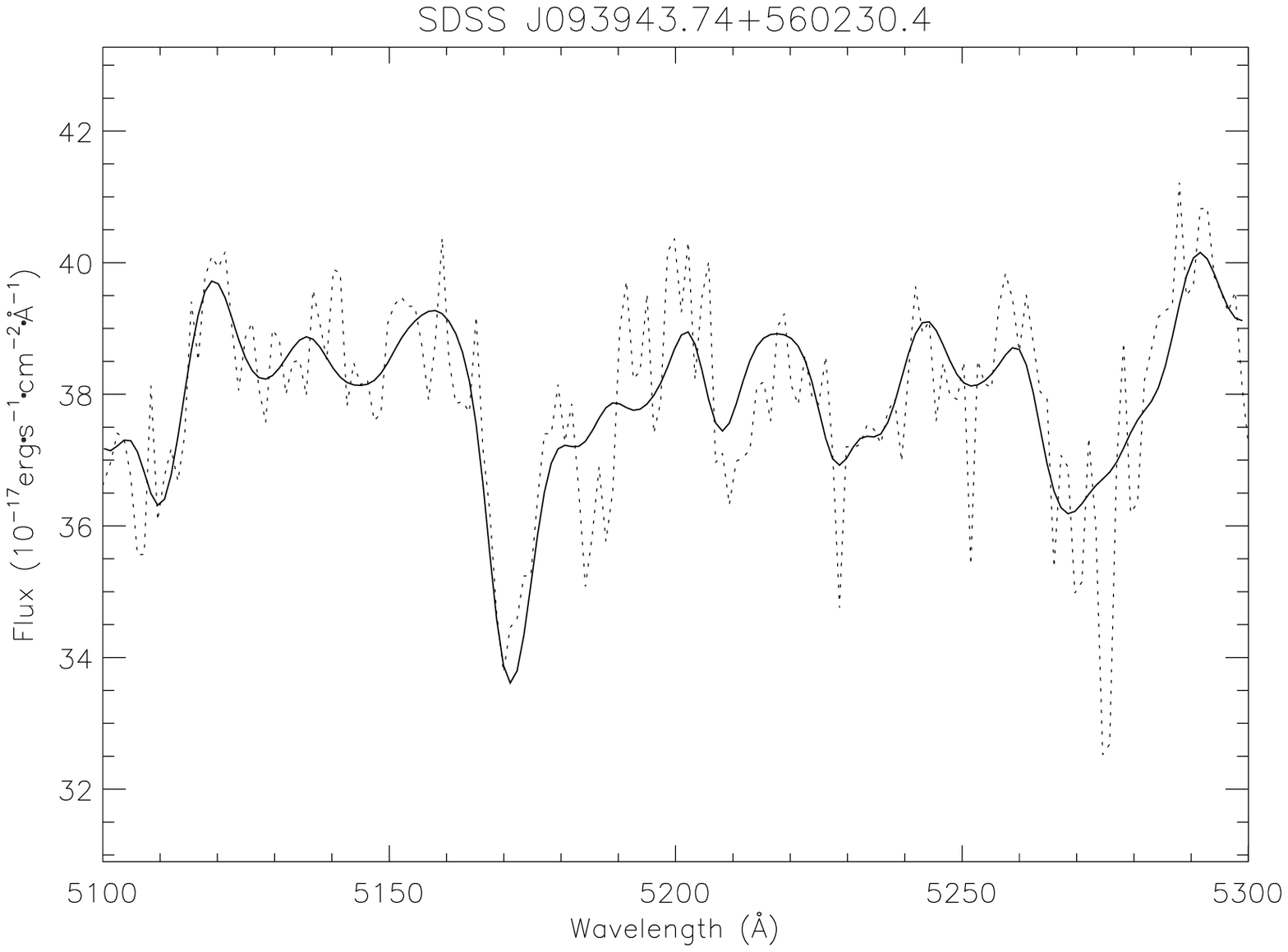}
\centering\includegraphics[height = 5cm,width = 8cm]{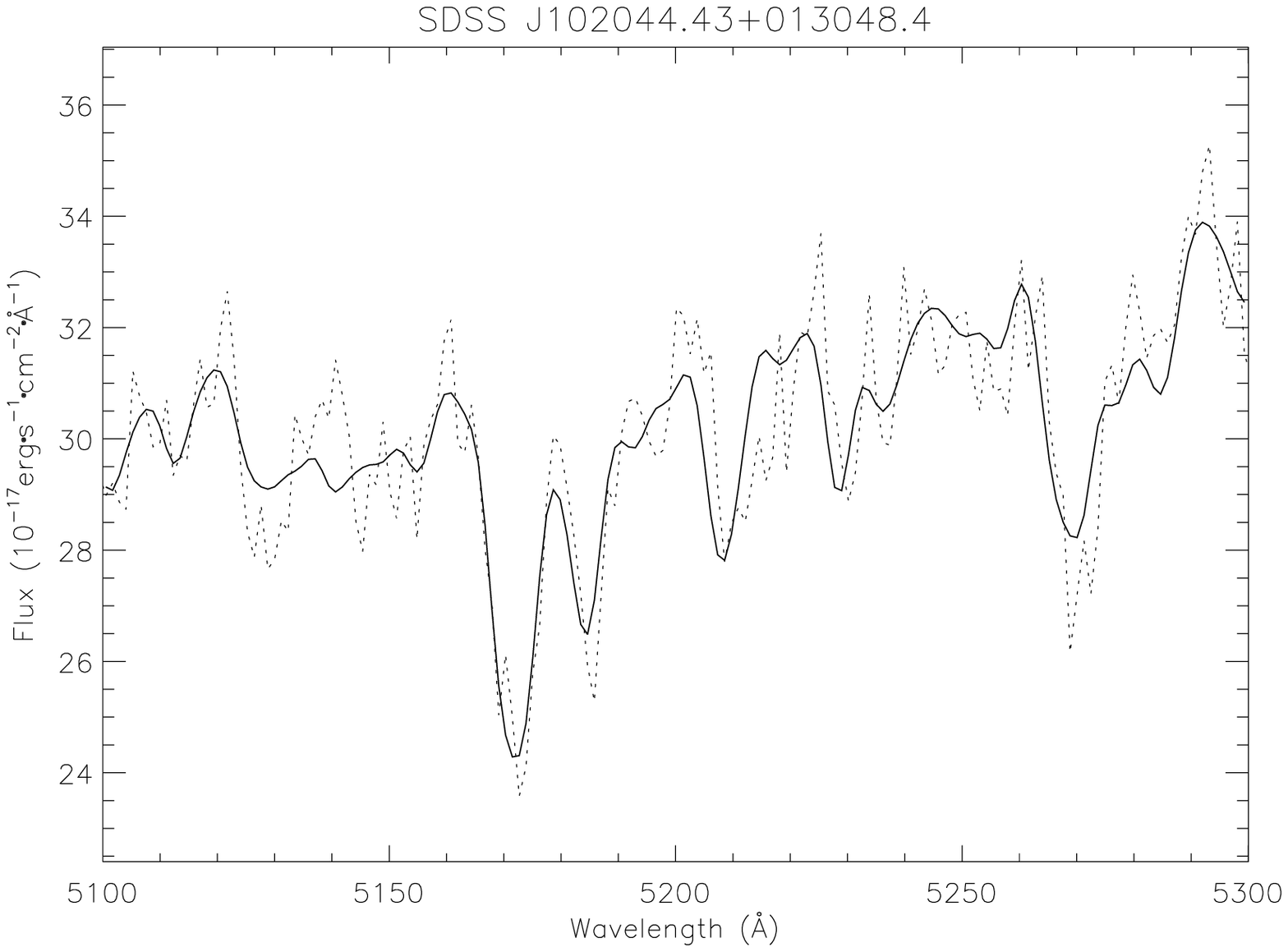}
\centering\includegraphics[height = 5cm,width = 8cm]{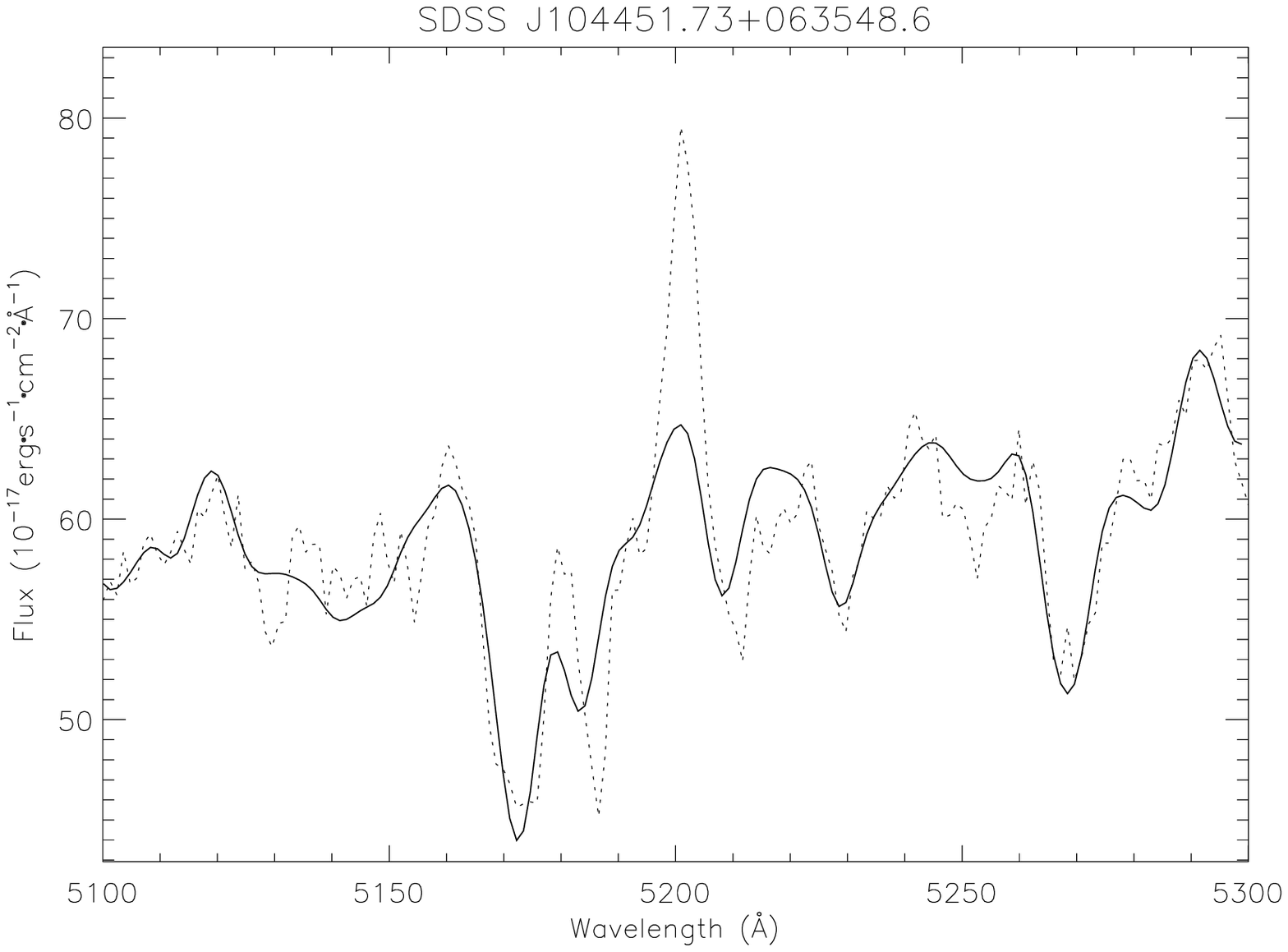}
\centering\includegraphics[height = 5cm,width = 8cm]{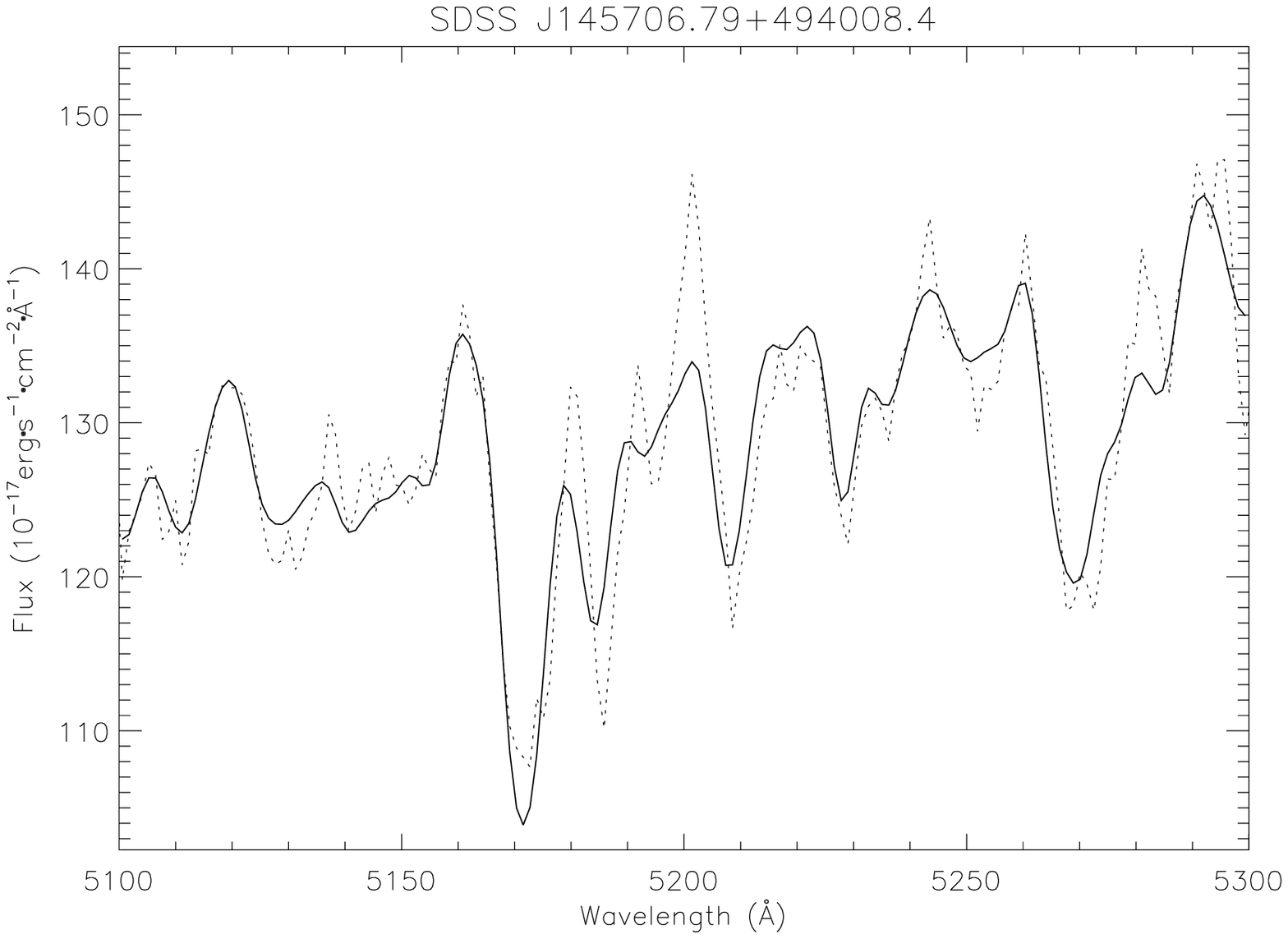}
\centering\includegraphics[height = 5cm,width = 8cm]{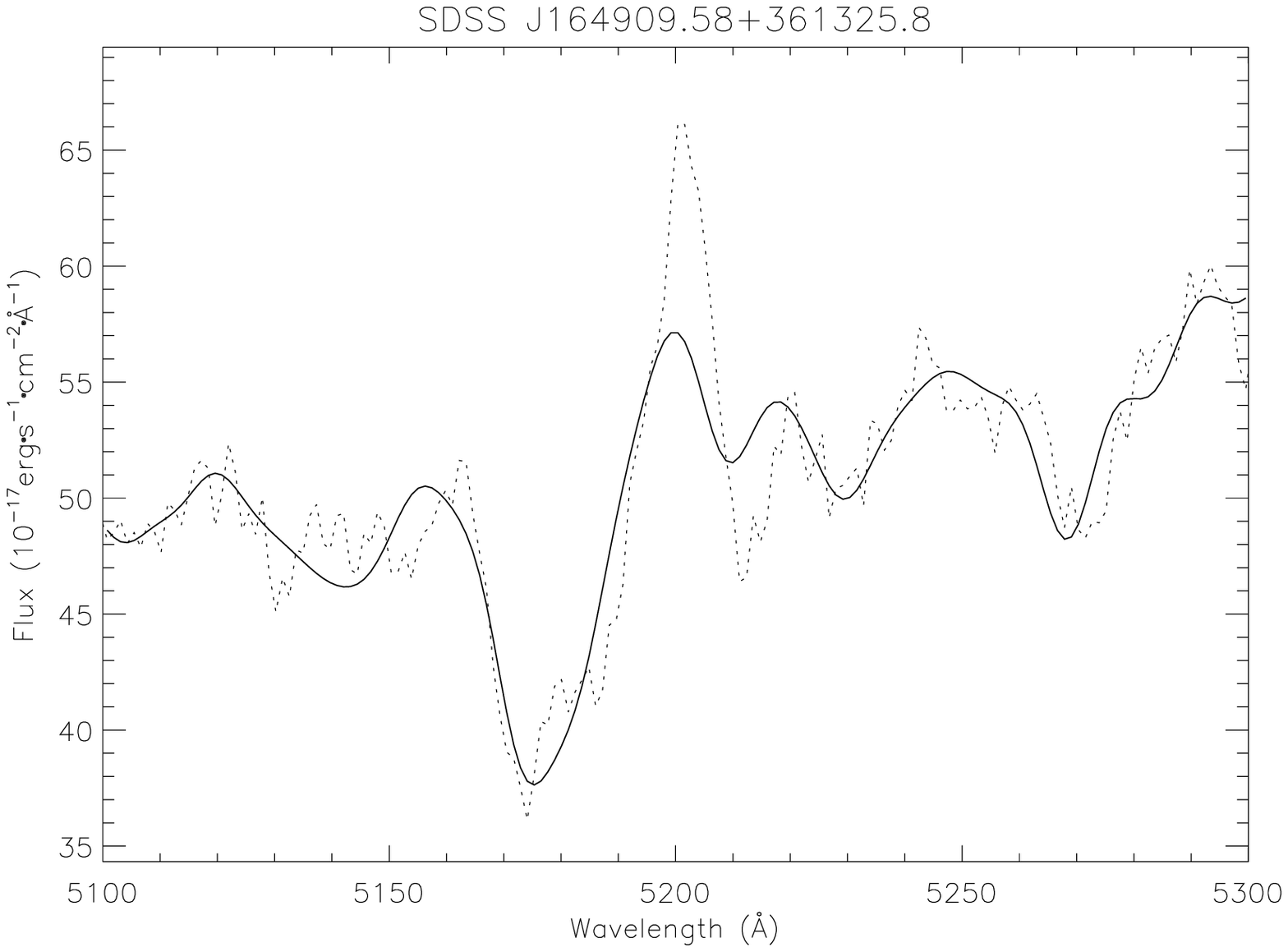}
\centering\includegraphics[height = 5cm,width = 8cm]{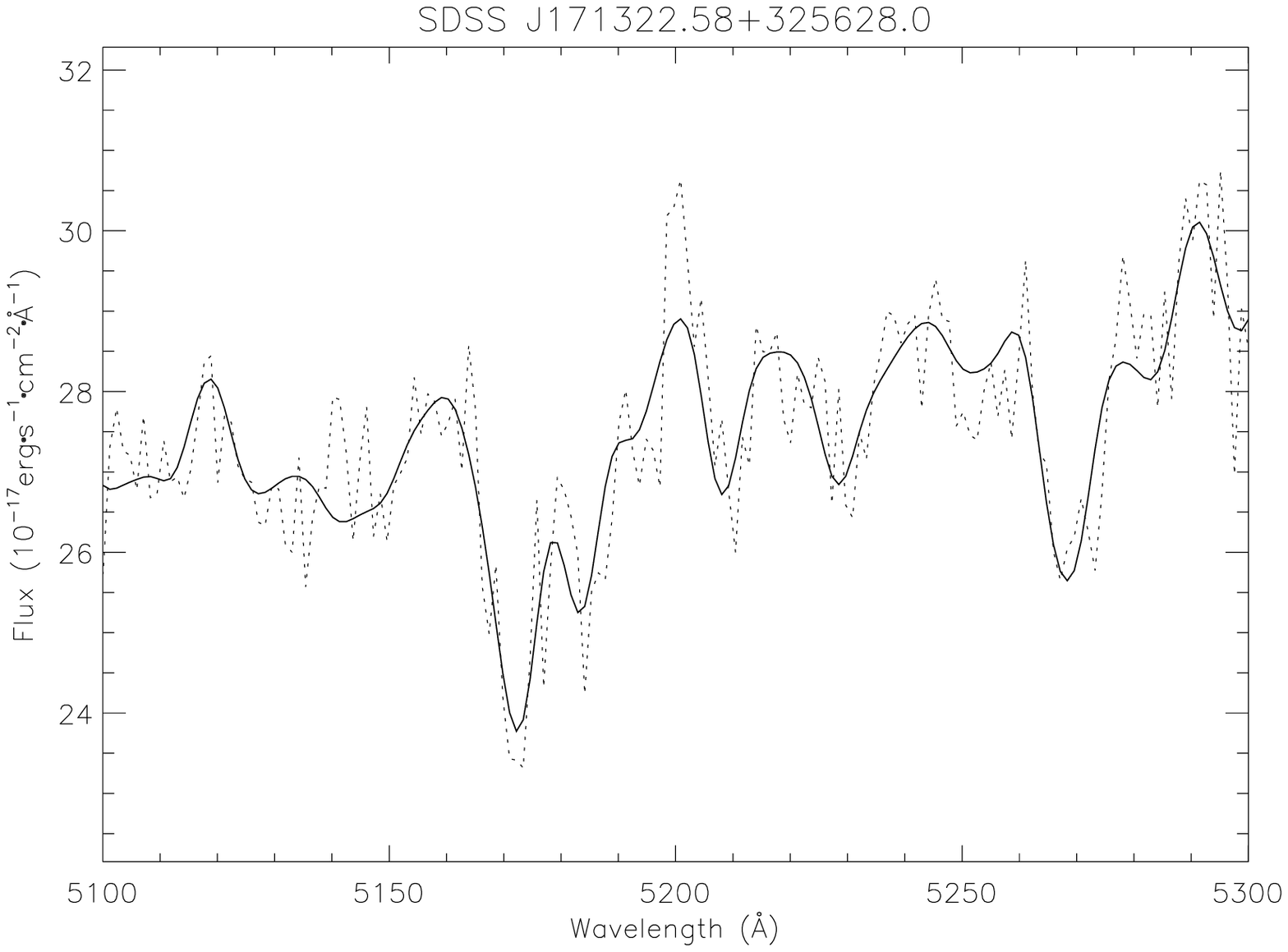}
\centering\includegraphics[height = 5cm,width = 8cm]{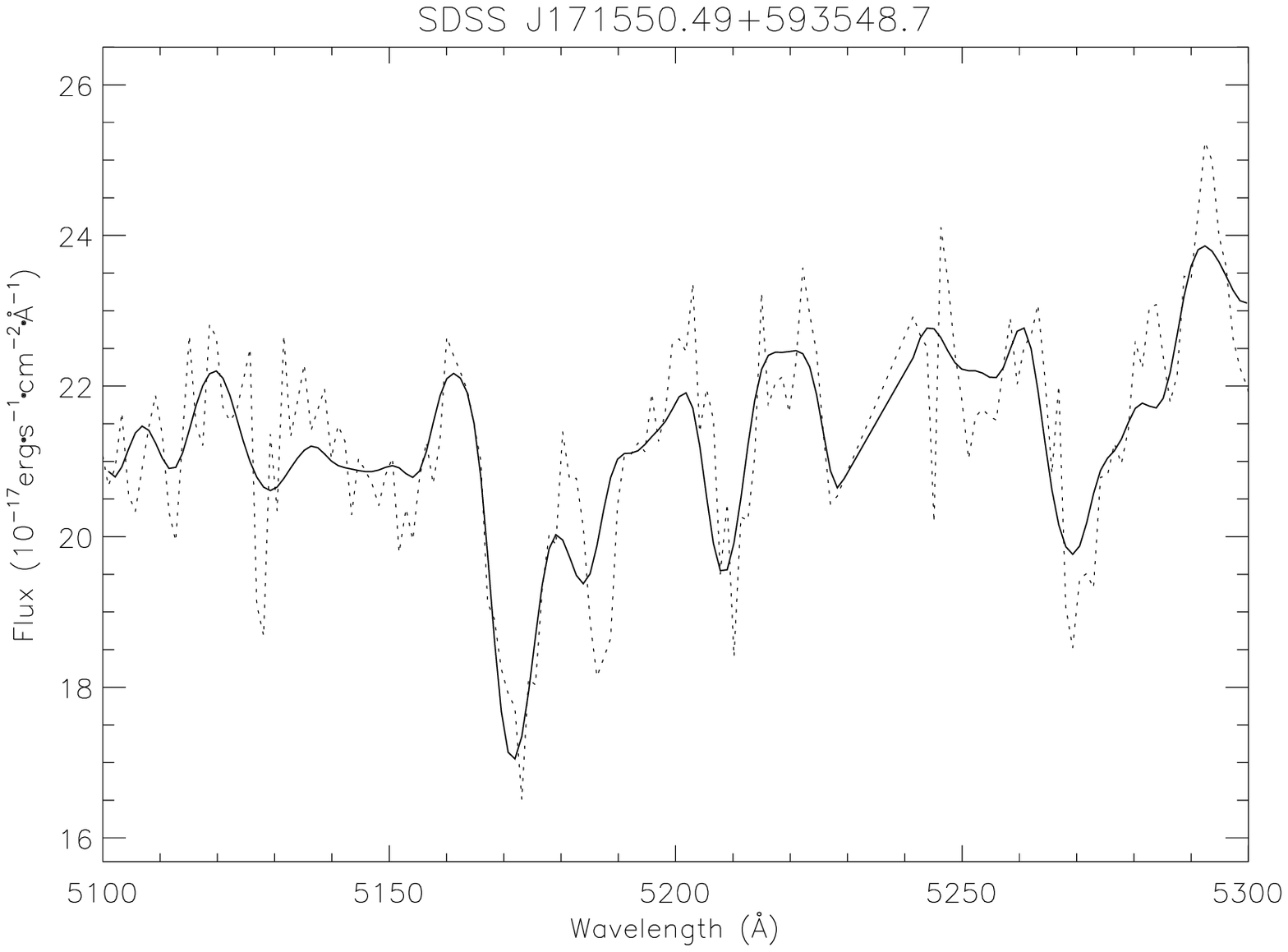}
\caption{The best fitted results for absorption features near
MgI$\lambda5175\AA$. Dotted line represents the observed spectrum, solid line
is for the best fitted results.
}
\label{abs}
\end{figure*}

\begin{figure*}
\centering\includegraphics[height = 8cm,width = 10cm]{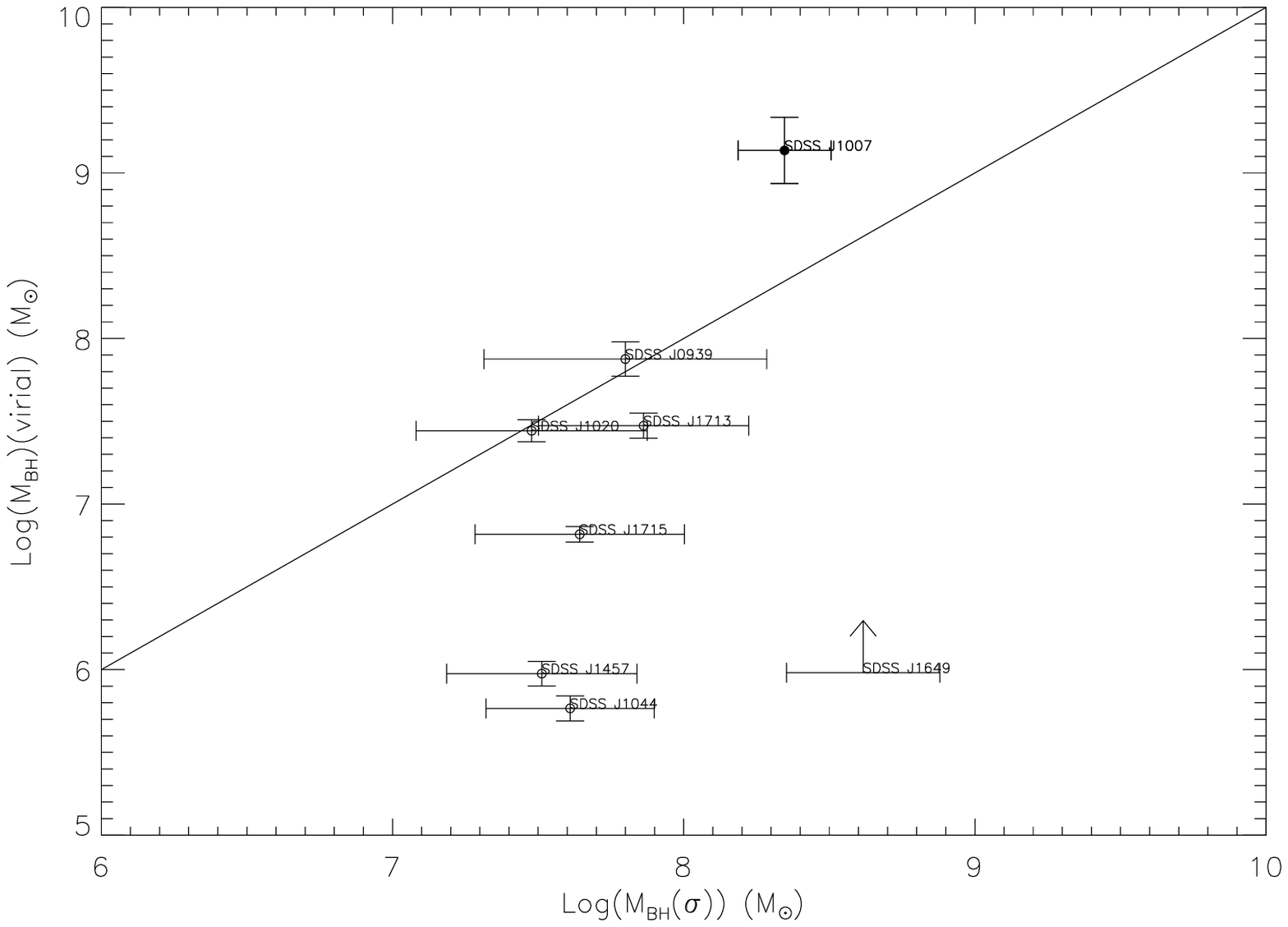}
\caption{The correlation between two kinds of BH masses, $M_{BH}(\sigma)$
is estimated through stellar velocity dispersion, $M_{BH}(virial)$ is estimated
by line width of broad H$\alpha$ and continuum luminosity. Solid circle
represents the object of SDSS J1007, of which the stellar velocity dispersion
is substituted by line width of narrow emission lines. The upward arrow for object SDSS J1649 indicates that actual virial BH mass should be larger than the one.
Solid line represents $M_{BH}(\sigma) = M_{BH}(virial)$.
}
\label{mass2}
\end{figure*}

\begin{figure*}
\centering\includegraphics[height = 8cm,width = 10cm]{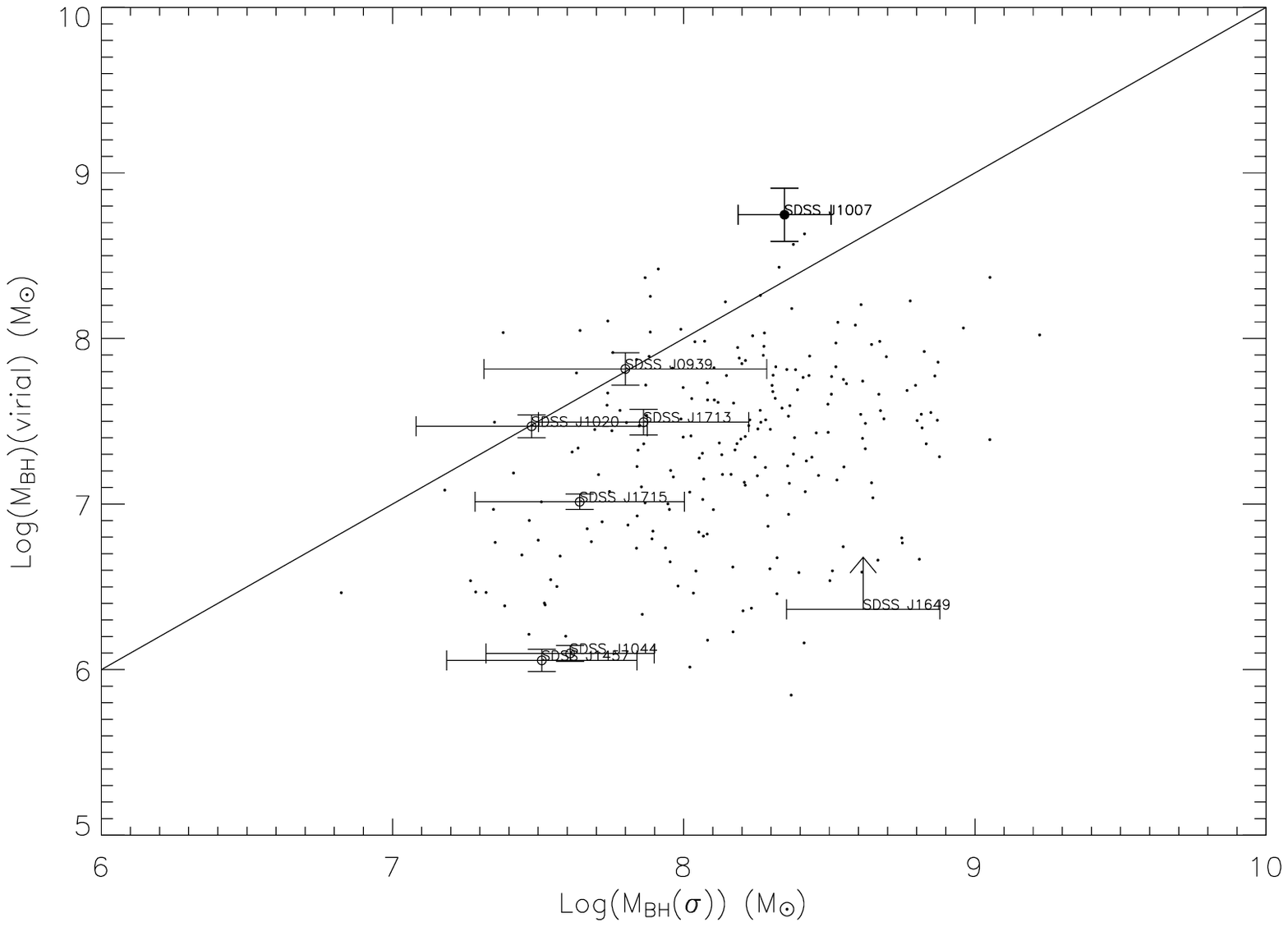}
\caption{The correlation between two kinds of BH masses after the correction of internal reddening effects, $M_{BH}(\sigma)$
is estimated through stellar velocity dispersion, $M_{BH}(virial)$ is estimated
by line width of broad H$\alpha$ and continuum luminosity. Solid circle
represents the object of SDSS J1007, of which the stellar velocity dispersion
is substituted by line width of narrow emission lines. The upward arrow for object SDSS J1649 indicates that actual virial BH mass should be larger than the one.
Solid line represents $M_{BH}(\sigma) = M_{BH}(virial)$. 
The solid small dots are the objects selected from DR6 with shifted velocities 
larger than 180${\rm km/s}$.
}
\label{mass2n}
\end{figure*}

\begin{figure*}
\centering\includegraphics[height = 8cm,width = 10cm]{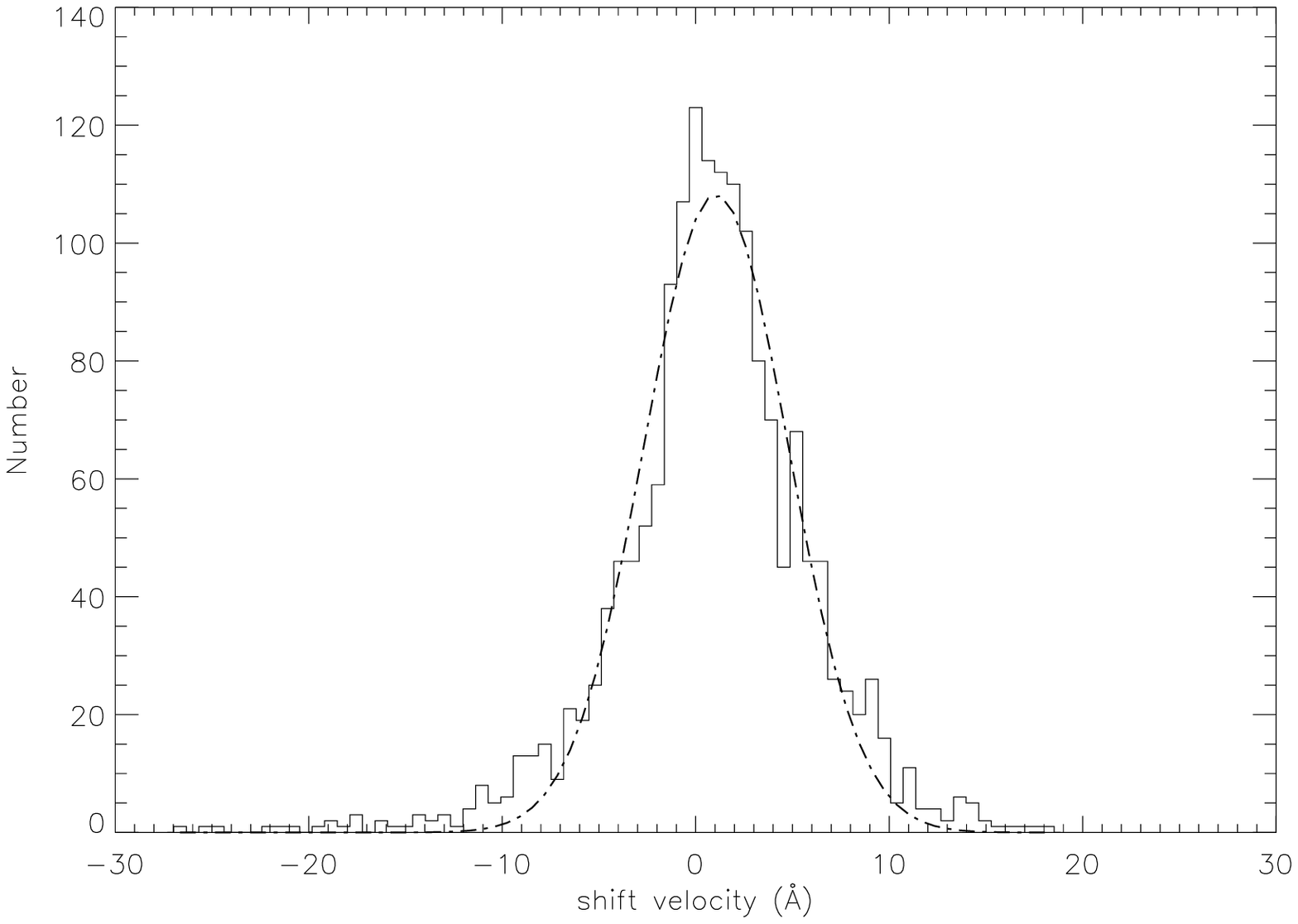}
\caption{The distributions of shifted velocities of the selected 1672 
AGN with standard gaussian broad H$\alpha$. Thin line represents the 
histogram of the shift velocities, thick dot-dashed line represents the 
fitted results by a gaussian function.
}
\label{nrat}
\end{figure*}

\end{document}